\makeatletter
\long\def\@makecaption#1#2{\ifx\@captype\@IEEEtablestring%
\footnotesize\begin{center}{\normalfont\footnotesize #1}\\
{\normalfont\footnotesize\scshape #2}\end{center}%
\@IEEEtablecaptionsepspace
\else
\@IEEEfigurecaptionsepspace
\setbox\@tempboxa\hbox{\normalfont\footnotesize {#1.}~~ #2}%
\ifdim \wd\@tempboxa >\hsize%
\setbox\@tempboxa\hbox{\normalfont\footnotesize {#1.}~~ }%
\parbox[t]{\hsize}{\normalfont\footnotesize \noindent\unhbox\@tempboxa#2}%
\else
\hbox to\hsize{\normalfont\footnotesize\hfil\box\@tempboxa\hfil}\fi\fi}
\makeatother

\documentclass{sig-alternate_tight}

\usepackage{subfigure}
\usepackage{graphicx}
\usepackage{amsmath}
\usepackage{bm}
\usepackage{url}
\usepackage{stmaryrd}
\usepackage{balance}
\usepackage{amssymb}
\usepackage{pgfplots}
\usepackage{pifont}
\usepackage[usenames,dvipsnames]{pstricks}
\usepackage{epsfig}
\usepackage{booktabs}
\usepackage{pgffor}
\usepackage{tikz}
\usepackage{multirow}
\usepackage{amsmath}

\newcommand{\mb}{\mathbf}

\newcommand{\our}{\textsc{MCD}}
\newcommand{\ourkmeans}{\textsc{Kmeans}}
\newcommand{\ourcut}{\textsc{Ncut}}
\newcommand{\ourinfluence}{\textsc{SIclus}}

\usepackage{algorithm}
\usepackage{algorithmic}

\begin{document}
\title{Mutual Community Detection across Multiple Partially Aligned Social Networks}

\numberofauthors{2}
\author{
\alignauthor
Jiawei~Zhang\\
      \affaddr{University of Illinois at Chicago}\\
      \affaddr{Chicago, IL, USA}\\
       \email{jzhan9@uic.edu}
\alignauthor
Philip~S.~Yu\\
      \affaddr{University of Illinois at Chicago, \\Chicago, IL, USA}\\
      \affaddr{Institute for Data Science, \\Tsinghua University, China}\\
       \email{psyu@cs.uic.edu}
}

\maketitle

\begin{abstract}

Community detection in online social networks has been a hot research topic in recent years. Meanwhile, to enjoy more social network services, users nowadays are usually involved in multiple online social networks simultaneously, some of which can share common information and structures. Networks that involve some common users are named as multiple ``partially aligned networks''. In this paper, we want to detect communities of multiple partially aligned networks simultaneously, which is formally defined as the ``Mutual Clustering'' problem. The ``Mutual Clustering'' problem is very challenging as it has two important issues to address: (1) how to preserve the network characteristics in mutual community detection? and (2) how to utilize the information in other aligned networks to refine and disambiguate the community structures of the shared users? To solve these two challenges, a novel community detection method, {\our} (Mutual Community Detector), is proposed in this paper. {\our} can detect social community structures of users in multiple partially aligned networks at the same time with full considerations of (1) characteristics of each network, and (2) information of the shared users across aligned networks. Extensive experiments conducted on two real-world partially aligned heterogeneous social networks demonstrate that {\our} can solve the ``Mutual Clustering'' problem very well.

\end{abstract}



\section{Introduction}\label{sec:intro}

Nowadays, online social networks which can provide users with various services have become ubiquitous in our daily life. The services provided by social networks are very diverse, e.g., make new friends online, read and write comments on recent news, recommend products and locations, etc. Real-world social networks which can provide these services usually have heterogeneous information, involving various kinds of information entities (e.g., users, locations, posts) and complex connections (e.g., social links among users, purchase links between users and products). Meanwhile, among these services provided by social networks, community detection techniques play a very important role. For example, organizing online friends into different categories (e.g., ``family members'', ``celebrities'', and ``classmates'') in Facebook and group-level recommendations of products in e-commerce sites are all based on community structures of users detected from the networks. 


Meanwhile, as proposed in \cite{KZY13, ZKY13, ZKY14, ZYZ14}, to enjoy more social services, users nowadays are usually involved in multiple online social networks simultaneously, e.g., Facebook, Twitter and Foursquare. Furthermore, some of these networks can share common information either due to the common network establishing purpose or because of similar network features \cite{ZY15}. Across these networks, the common users are defined as the \textit{anchor users}, while the remaining non-shared users are named as the \textit{non-anchor users}. Connections between \textit{anchor users'} accounts in different networks are defined as the \textit{anchor links}. The networks partially aligned by \textit{anchor links} are called \textit{multiple partially aligned networks}.

In this paper, we want to detect the communities of each network across multiple \textit{partially aligned social networks} simultaneously, which is formally defined as the \textit{Mutual Clustering} problem. The goal is to distill relevant information from another social network to compliment knowledge directly derivable from each network to improve the clustering or community detection, while preserving the distinct characteristics of each individual network. The \textit{Mutual Clustering} problem is very important for online social networks and can be the prerequisite for many concrete social network applications: (1) \textit{network partition}: Detected communities can usually represent small-sized subgraphs of the network, and (2) \textit{comprehensive understanding of user social behaviors}: Community structures of the shared users in multiple aligned networks can provide a complementary understanding of their social interactions in online social networks.


Besides its importance, the \textit{Mutual Clustering} problem is a novel problem and different from existing clustering problems, including: (1) \textit{consensus clustering}, \cite{GF08, LDJ07, NC07, LBRFFP13, LD13} which aims at achieving a consensus result of several input clustering results about the same data; (2) \textit{multi-view clustering}, \cite{BS04, CNH13} whose target is to partition objects into clusters based on their different representations, e.g., clustering webpages with text information and hyperlinks; (3) \textit{multi-relational clustering}, \cite{YHY07, BG05} which focuses on clustering objects in one relation (called target relation) using information in multiple inter-linked relations; and (4) \textit{co-regularized multi-domain graph clustering} \cite{CZGW13}, which relaxes the \textit{one-to-one} constraints on node correspondence relationships between different views in multi-view clustering to ``\textit{uncertain}'' mappings. In \cite{CZGW13}, prior knowledge about the weights of mappings is required and each view is actually a homogeneous network (more differences are summarized in Section~\ref{sec:relatedwork}). Unlike these existing clustering problems, the \textit{Mutual Clustering} problem aims at detecting the communities for multiple networks involving both anchor and non-anchor users simultaneously and each network contains heterogeneous information about users' social activities. A more detailed comparison of  \textit{Mutual Clustering} problem with these related problems is available in Table~\ref{tab:related}.

\begin{table*}[t]
\scriptsize
\centering
{
\caption{Summary of related problems.}\label{tab:related}
\begin{tabular}{|l|c|c|c|c|c|}
\hline
& \textbf{Mutual}
& \textbf{Co-Regularized}
& \textbf{Consensus}
& \textbf{Multi-View}
& \textbf{Multi-Relational} \\

& \textbf{Clustering}
& \textbf{Graph}
& \textbf{Clustering}
& \textbf{Clustering}
& \textbf{Clustering} \\
Property
& \textbf{}
& \textbf{Clustering \cite{CZGWSW13}}
& \textbf{\cite{LBRFFP13, LD13, ZL11-2}}
& \textbf{\cite{BS04, CNH13, CKLS09}}
& \textbf{\cite{YHY07, BG05}} \\
\hline

input
&multiple networks
&data with multi-views
&multiple clusters
&data with multi-views
&data of multi-relations\\


output
&clusters of
&clusters of items
&consensus of
&clusters of items
&clusters of item in\\

&each network
&in each view
&input clusters
&across views
&the target relation\\


\# networks
&multiple
&multiple
&n/a
&multiple
&single\\


network type
&heterogeneous
&homogeneous
&n/a
&homogeneous
&heterogeneous\\


network aligned?	
&partially aligned
&no
&n/a
&no
&no\\

network connections
&anchor links
&uncertain mappings
&n/a
&certain mappings
&n/a\\

\hline

\end{tabular}
}
\end{table*}




Despite its importance and novelty, the \textit{Mutual Clustering} is very challenging to solve due to:
\begin{itemize}

\item \textit{Closeness Measure}: Users in heterogeneous social networks can be connected with each other by various direct and indirect connections. A general closeness measure among users with such connection information is the prerequisite for addressing the \textit{mutual clustering} problem.


\item \textit{Network Characteristics}: Social networks usually have their own characteristics, which can be reflected in the community structures formed by users. Preservation of each network's characteristics (i.e., some unique structures in each network's detected communities) is very important in the \textit{Mutual Clustering} problem.

\item \textit{Mutual Community Detection}: Information in different networks can provide us with a more comprehensive understanding about the anchor users' social structures. For anchor users whose community structures are not clear based on in formation in one network, utilizing the heterogeneous information in aligned networks to refine and disambiguate the community structures about the anchor users. However, how to achieve such a goal is still an open problem.

\item \textit{Lack of Metrics}. \textit{Mutual Clustering} problem is a new problem and few existing metrics can be applied to evaluate the comprehensive performance of \textit{Mutual Clustering} methods.



\end{itemize}

To solve all these challenges, a novel cross-network community detection method, {\our} (Mutual Community Detector), is proposed in this paper. {\our} maps the complex relationships in the social network into a heterogeneous information network \cite{SHYYW11} and introduces a novel meta-path based closeness measure, \textit{HNMP-Sim}, to utilize both direct and indirect connections among users in closeness scores calculation. With full considerations of the network characteristics, {\our} exploits the information in aligned networks to refine and disambiguate the community structures of the multiple networks concurrently. Besides traditional \textit{quality} and \textit{consensus} metrics, we define a novel general metric, $IQC$ (Integrated Quality \& Consensus), to evaluate the performance of mutual clustering methods.

This paper is organized as follows: In Section~\ref{sec:formulation}, we formulate the problem. Section~\ref{sec:method} introduces the \textit{Mutual Clustering} methods. Section~\ref{sec:experiment} shows the experiment results. In Sections~\ref{sec:relatedwork} and \ref{sec:conclusion}, we give the related works and conclude this paper.



\section{Problem Formulation}
\label{sec:formulation}

\begin{figure}[t]
\centering
    \begin{minipage}[l]{1.0\columnwidth}
      \centering
      \includegraphics[width=1.0\textwidth]{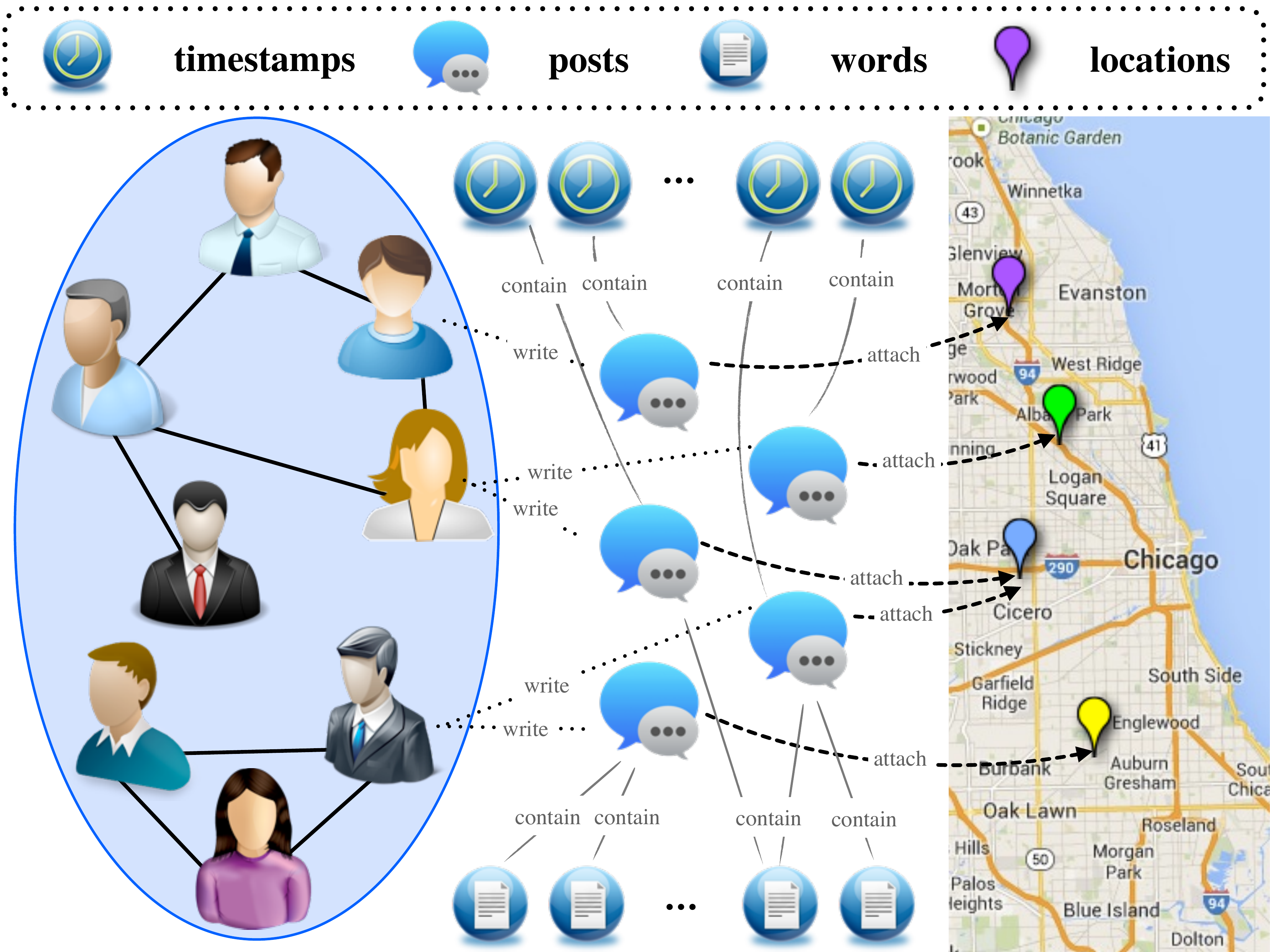}
    \end{minipage}
\caption{Heterogeneous online social networks.}\label{eg_fig3}
\end{figure}

The networks studied in this paper are Foursquare and Twitter. Users in both Foursquare and Twitter can follow other users, write tips/tweets, which can contain timestamps, text content and location check-ins. As a result, both Foursquare and Twitter can be modeled as heterogeneous information networks $G = (V, E)$, where $V = \mathcal{U} \cup \mathcal{P} \cup \mathcal{L} \cup \mathcal{T} \cup \mathcal{W}$ is the set of different types of nodes in the network and $\mathcal{U}$,  $\mathcal{P}$,  $\mathcal{L}$,  $\mathcal{T}$,  $\mathcal{W}$ are the node sets of users, posts, location check-ins, timestamps and words respectively, while $E = \mathcal{E}_{s} \cup \mathcal{E}_{p} \cup \mathcal{E}_{l} \cup \mathcal{E}_{t} \cup \mathcal{E}_{w}$ is set of directed links in the network and $\mathcal{E}_{s}$, $\mathcal{E}_{p}$, $\mathcal{E}_{l}$, $\mathcal{E}_{t}$ and $\mathcal{E}_{w}$ are the sets of social links among users, links between users and posts and those between posts and location-checkins, timestamps as well as words respectively. To illustrate the structure of the heterogeneous network studied in this paper, we also give an example in Figure~\ref{eg_fig3}. As shown in the figure, users in the network can be extensively connected with each other by different types of links (e.g., social links, co-location checkins connections).

The multiple aligned networks can be modeled as $\mathcal{G} = (G_{set}, A_{set})$, where $G_{set} = \{G^{(1)}, G^{(2)}, \ldots, G^{(n)}\}, |G_{set}| = n$ is the set of $n$ heterogeneous information networks and $A_{set} = \{A^{(1,2)}, \ldots, A^{(1,n)}, A^{(2,3)}, \ldots, A^{((n - 1),n)}\}$ is the set of undirected anchor links between different heterogeneous networks in $G_{set}$. In this paper, we will follow the definitions about ``anchor user'', ``non-anchor user'', ``anchor link'', etc. proposed in \cite{KZY13, ZKY13, ZKY14, ZYZ14} and the constraint on anchor links is ``\textit{one-to-one}'', i.e., each user can have one account in on network. The case that users have multiple accounts in online social networks can be resolved with method introduced in \cite{TZ14}, where these duplicated accounts can be aggregated in advance to form one unique vitural account in advance and the anchor links connecting these vitural accounts will be still ``\textit{one-to-one}''. Different from \cite{KZY13, ZKY13}, networks studied in this paper are all \textit{partially aligned} \cite{ZKY14, ZYZ14}.




\noindent \textbf{Mutual Clustering Problem}: For the given multiple aligned heterogeneous networks $\mathcal{G}$, the \textit{Mutual Clustering} problem aims to obtain the optimal communities $\{\mathcal{C}^{(1)}, \mathcal{C}^{(2)}, \cdots, \mathcal{C}^{(n)}\}$ for $\{G^{(1)}, G^{(2)}, \cdots, G^{(n)}\}$ simultaneously, where\\ $\mathcal{C}^{(i)} = \{U^{(i)}_1, U^{(i)}_2, \ldots, U^{(i)}_{k^{(i)}}\}$ is a partition of the users set $\mathcal{U}^{(i)}$ in $G^{(i)}$, $k^{(i)} = \left | \mathcal{C}^{(i)} \right |$, $U^{(i)}_l \cap U^{(i)}_m = \emptyset$, $\forall\ l,m \in \{1, 2, \ldots, k^{(i)}\}$ and $\bigcup_{j = 1}^{k^{(i)}} U^{(i)}_j = \mathcal{U}^{(i)}$. Users in each detected cluster are more densely connected with each other than with users in other clusters. In this paper, we focus on studying the hard (i.e., non-overlapping) clustering of users in online social networks.


\section{Proposed Methods}
\label{sec:method}


\begin{table*}[t]
\scriptsize
\centering
{
\caption{Summary of HNMPs.}\label{tab:meta_path}
\begin{tabular}{llll}
\hline
\textbf{ID}
&\textbf{Notation}
& \textbf{Heterogeneous Network Meta Path}
& \textbf{Semantics}\\
\hline
\hline

1
&U $\to$ U
&User $\xrightarrow{follow}$ User
&Follow\\

2
&U $\to$ U $\to$ U
&User $\xrightarrow{follow}$ User $\xrightarrow{follow}$ User
&Follower of Follower\\

3
&U $\to$ U $\gets$ U
&User $\xrightarrow{follow}$ User $\xrightarrow{follow^{-1}}$ User
&Common Out Neighbor\\

4
&U $\gets$ U $\to$ U
&User $\xrightarrow{follow^{-1}}$ User $\xrightarrow{follow}$ User
&Common In Neighbor\\

\hline


5
&U $\to$ P $\to$ W $\gets$ P $\gets$ U
&User $\xrightarrow{write}$ Post $\xrightarrow{contain}$ Word 
&Posts Containing Common Words\\

&&\ \ \ \ \ \ \ \ \ \ \ \ \ \ \ \ \ \ \ \ $\xrightarrow{contain^{-1}}$ Post $\xrightarrow{write^{-1}}$ User&\\

6
&U $\to$ P $\to$ T $\gets$ P $\gets$ U
&User $\xrightarrow{write}$ Post $\xrightarrow{contain}$ Time 
&Posts Containing Common Timestamps\\

&&\ \ \ \ \ \ \ \ \ \ \ \ \ \ \ \ \ \ \ \ $\xrightarrow{contain^{-1}}$ Post $\xrightarrow{write^{-1}}$ User&\\

7
&U $\to$ P $\to$ L $\gets$ P $\gets$ U
&User $\xrightarrow{write}$ Post $\xrightarrow{attach}$ Location
&Posts Attaching Common Location Check-ins\\

&&\ \ \ \ \ \ \ \ \ \ \ \ \ \ \ \ \ \ \ \ $\xrightarrow{attach^{-1}}$ Post $\xrightarrow{write^{-1}}$ User&\\
\hline

\end{tabular}
}
\end{table*}

A co-regularization based multi-view clustering model was proposed in \cite{CZGW13}, which achieves the clustering results of nodes across multi-view by minimizing absolute clustering disagreement of all nodes (both shared and non-shared nodes). It cannot be applied to address the \textit{Mutual Clustering} problem, as in {mutual clustering}, we only exploit information across networks to refine the social community structures of anchor users only, while non-anchor users social community structures are not affected and can preserve their characteristics. To solve the \textit{Mutual Clustering} problem, a novel community detection method, {\our}, will be proposed in this section. By mapping the social network relations into a heterogeneous information network, we use the concept of social meta path to define closeness measure among users in Section  3.1. Based on this similarity measure, we introduce the network characteristics preservation independent clustering method in Section 3.2 and normalized discrepancy based co-clustering method in Section 3.3. To preserve network characteristics and use information in other networks to refine community structures mutually at the same time, we study the mutual clustering problem in Section 3.4.



\subsection{HNMP-Sim}

Many existing similarity measures, e.g., ``Common Neighbor'' \cite{HZ11}, ``Jaccard's Coefficient'' \cite{HZ11}, defined for homogeneous networks cannot capture all the connections among users in heterogeneous networks. To use both direct and indirect connections among users in calculating the similarity score among users in the heterogeneous information network, we introduce meta path based similarity measure HNMP-Sim in this section.


\subsubsection{Meta Paths in Heterogeneous Networks}

In heterogeneous networks, pairs of nodes can be connected by different paths, which are sequences of links in the network. Meta paths \cite{SHYYW11, SYH09} in heterogeneous networks, i.e., \textit{heterogeneous network meta paths} (HNMPs), can capture both direct and indirect connections among nodes in a network. The length of a meta path is defined as the number of links that constitute it. Meta paths in networks can start and end with various node types. However, in this paper, we are mainly concerned about those starting and ending with users, which are formally defined as the \textit{social HNMPs}. The notation, definition and semantics of $7$ different \textit{social HNMPs} used in this paper are listed in Table~\ref{tab:meta_path}. To extract the social meta paths, prior domain knowledge about the network structure is required.

\subsubsection{HNMP-based Similarity}

These $7$ different social HNMPs in Table~\ref{tab:meta_path} can cover lots of connections among users in networks. Some meta path based similarity measures have been proposed so far, e.g., the \textit{PathSim} proposed in \cite{SHYYW11}, which is defined for undirected networks and considers different meta paths to be of the same importance. To measure the social closeness among users in directed heterogeneous information networks, we extend \textit{PathSim} to propose a new closeness measure as follows.



%


\noindent \textbf{Definition 1} (HNMP-Sim): Let $\mathcal{P}_i(x \rightsquigarrow y)$ and $\mathcal{P}_i(x \rightsquigarrow \cdot)$ be the sets of path instances of HNMP \# $i$ going from $x$ to $y$ and those going from $x$ to other nodes in the network. The HNMP-Sim (HNMP based Similarity) of node pair $(x, y)$ is defined as
$$\mbox{HNMP-Sim}(x, y) = \sum_i \omega_i \left(\frac{\left| \mathcal{P}_i(x \rightsquigarrow y) \right| + \left| \mathcal{P}_i(y \rightsquigarrow x) \right|} {{\left| \mathcal{P}_i(x \rightsquigarrow \cdot) \right| } + \left| \mathcal{P}_i(y \rightsquigarrow \cdot) \right|}\right),$$
where $\omega_i$ is the weight of HNMP \# $i$ and $\sum_i \omega_i = 1$. In this paper, the weights of different HNMPs can be automatically adjusted by applying the technique proposed in \cite{ZY15}. 


Let $\mathbf{A}_i$ be the \textit{adjacency matrix} corresponding to the HNMP \# i among users in the network and $\mb{A}_i(m, n) = k$ iff there exist $k$ different path instances of HNMP \# $i$ from user $m$ to $n$ in the network. Furthermore, the similarity score matrix among users of HNMP \# $i$ can be represented as $\mb{S}_i = \left( \mb{D}_i + \bar{\mb{D}}_i \right)^{-1} \left( \mb{A}_i + \mb{A}_i^T \right)$, where $\mb{A}_i^T$ denotes the transpose of $\mb{A}_i$, diagonal matrices $\mb{D}_i$ and $\bar{\mb{D}}_i$ have values $\mb{D}_i{(l,l)} = \sum_m \mb{A}_i{(l,m)}$ and $\bar{\mb{D}}_i{(l,l)} = \sum_m (\mb{A}_i^T){(l,m)}$ on their diagonals respectively. The HNMP-Sim matrix of the network which can capture all possible connections among users is represented as follows:
$$\mathbf{S} = \sum_i \omega_i \mb{S}_i = \sum_i \omega_i \left( \left( \mb{D}_i + \bar{\mb{D}}_i \right)^{-1} \left( \mb{A}_i + \mb{A}_i^T \right) \right).$$




\subsection{Network Characteristic Preservation Clustering}
Clustering each network independently can preserve each networks characteristics effectively as no information from external networks will interfere with the clustering results. Partitioning users of a certain network into several clusters will cut connections in the network and lead to some costs inevitably. Optimal clustering results can be achieved by minimizing the clustering costs.


For a given network $G$, let $\mathcal{C} = \{U_1, U_2, \ldots, U_k\}$ be the community structures detected from $G$. Term $\overline{U_i} = \mathcal{U} - U_i$ is defined to be the complement of set $U_i$ in $G$. Various cost measure of partition $\mathcal{C}$ can be used, e.g., \textit{cut} \cite{WL93} and \textit{normalized cut} \cite{SM00}:
$$cut(\mathcal{C}) =  \frac{1}{2}\sum_{i = 1}^k S(U_i, \overline{U_i}) = \frac{1}{2}\sum_{i = 1}^k \sum_{u \in U_i, v \in \overline{U_i}} S(u, v),$$
$$Ncut(\mathcal{C}) = \frac{1}{2}\sum_{i = 1}^k \frac{S(U_i, \overline{U_i})}{S(U_i, \cdot)} = \sum_{i = 1}^k\frac{cut(U_i, \overline{U}_i)}{S(U_i, \cdot)},$$
where $S(u, v)$ denotes the HNMP-Sim between $u, v$ and $S(U_i, \cdot) = S(U_i, \mathcal{U}) = S(U_i, U_i) + S(U_i, \overline{U}_i)$. 


%

For all users in $\mathcal{U}$, their clustering result can be represented in the \textit{result confidence matrix} $\mb{H}$, where $\bf{H} = [\bf{h}_1,$ $\bf{h}_2,$ $\ldots,$ $\bf{h}_n]^T$, $n = |\mathcal{U}|$, $\mb{h}_i = (h_{i,1}, h_{i,2}, \ldots, h_{i,k})$ and $h_{i,j}$ denotes the confidence that $u_i \in \mathcal{U}$ is in cluster $U_j \in \mathcal{C}$. The optimal $\bf{H}$ that can minimize the normalized-cut cost can be obtained by solving the following objective function \cite{L07}:
\begin{align*}
\min_{\mb{H}}\ \ &\mbox{Tr} (\mb{H}^T\mb{L}\mb{H}),\\
s.t. \ \ &\mb{H}^T\mb{D}\mb{H} = \mb{I}.
\end{align*}
where $\mb{L} = \mb{D} - \mb{S}$, diagonal matrix $\mb{D}$ has $\mb{D}(i,i) = \sum_j \mb{S}(i,j)$ on its diagonal, and $\mb{I}$ is an identity matrix.


%
%
%


\subsection{Discrepancy based Clustering of Multiple Networks}

Besides the shared information due to common network construction purposes and similar network features \cite{ZY15}, anchor users can also have unique information (e.g., social structures) across aligned networks, which can provide us with a more comprehensive knowledge about the community structures formed by these users. Meanwhile, by maximizing the consensus (i.e., minimizing the ``\textit{discrepancy}'') of the clustering results about the anchor users in multiple partially aligned networks, we refine the clustering results of the anchor users with information in other aligned networks mutually. We can represent the clustering results achieved in $G^{(1)}$ and $G^{(2)}$ as $\mathcal{C}^{(1)} = \{U^{(1)}_1, U^{(1)}_2,$ $\cdots,$ $U^{(1)}_{k^{(1)}}\}$ and $\mathcal{C}^{(2)} = \{U^{(2)}_1, U^{(2)}_2, \cdots, U^{(2)}_{k^{(2)}}\}$ respectively.


Let $u_i$ and $u_j$ be two anchor users in the network, whose accounts in $G^{(1)}$ and $G^{(2)}$ are $u^{(1)}_i$, $u^{(2)}_i$, $u^{(1)}_j$ and $u^{(2)}_j$ respectively. If users $u^{(1)}_i$ and $u^{(1)}_j$ are partitioned into the same cluster in $G^{(1)}$ but their corresponding accounts $u^{(2)}_i$ and $u^{(2)}_j$ are partitioned into different clusters in $G^{(2)}$, then it will lead to a \textit{discrepancy} between the clustering results of $u^{(1)}_i$, $u^{(2)}_i$, $u^{(1)}_j$ and $u^{(2)}_j$ in aligned networks $G^{(1)}$ and $G^{(2)}$.

\noindent \textbf{Definition 2} (Discrepancy): The discrepancy between the clustering results of $u_i$ and $u_j$ across aligned networks $G^{(1)}$ and $G^{(2)}$ is defined as the difference of confidence scores of $u_i$ and $u_j$ being partitioned in the same cluster across aligned networks. Considering that in the clustering results, the confidence scores of $u^{(1)}_i$ and $u^{(1)}_j$ ($u^{(2)}_i$ and $u^{(2)}_j$ ) being partitioned into $k^{(1)}$ ($k^{(2)}$) clusters can be represented as vectors $\mb{h}_i^{(1)}$ and $\mb{h}_j^{(1)}$ ($\mb{h}_i^{(2)}$ and $\mb{h}_j^{(2)}$) respectively, while the confidences that $u_i$ and $u_j$ are in the same cluster in $G^{(1)}$ and $G^{(2)}$ can be denoted as $\mb{h}_i^{(1)} (\mb{h}_j^{(1)})^T$ and $\mb{h}_i^{(2)} (\mb{h}_j^{(2)})^T$. Formally, the discrepancy of the clustering results about $u_i$ and $u_j$ is defined to be $d_{ij}(\mathcal{C}^{(1)}, \mathcal{C}^{(2)}) = \left(\mb{h}_i^{(1)} (\mb{h}_j^{(1)})^T - \mb{h}_i^{(2)} (\mb{h}_j^{(2)})^T\right )^2$ if $u_i, u_j$ are both anchor users; and $d_{ij}(\mathcal{C}^{(1)}, \mathcal{C}^{(2)}) = 0$ otherwise.
Furthermore, the discrepancy of $\mathcal{C}^{(1)}$ and $\mathcal{C}^{(2)}$ will be:
\begin{align*}
d(\mathcal{C}^{(1)}, \mathcal{C}^{(2)}) &= \sum_i^{n^{(1)}} \sum_j^{n^{(2)}} d_{ij}(\mathcal{C}^{(1)}, \mathcal{C}^{(2)}),
\end{align*}
where ${n^{(1)}} = |\mathcal{U}^{(1)}|$ and ${n^{(2)}} = |\mathcal{U}^{(2)}|$. In the definition, non-anchor users are not involved in the discrepancy calculation, which is totally different from the clustering disagreement function (all the nodes are included) introduced in \cite{CZGW13}

%


%

However, considering that $d(\mathcal{C}^{(1)}, \mathcal{C}^{(2)})$ is highly dependent on the number of anchor users and anchor links between $G^{(1)}$ and $G^{(2)}$, minimizing $d(\mathcal{C}^{(1)}, \mathcal{C}^{(2)})$ can favor highly consented clustering results when the anchor users are abundant but have no significant effects when the anchor users are very rare. To solve this problem, we propose to minimize the \textit{normalized discrepancy} instead, which significantly differs from the absolute clustering disagreement cost used in \cite{CZGW13}.


\noindent \textbf{Definition 3} (Normalized Discrepancy) The normalized discrepancy measure computes the  differences of clustering results in two aligned networks as a fraction of the discrepancy with regard to the number of anchor users across partially aligned networks:
$$Nd(\mathcal{C}^{(1)}, \mathcal{C}^{(2)}) = \frac{d(\mathcal{C}^{(1)}, \mathcal{C}^{(2)})}{\left( \left | A^{(1,2)} \right | \right) \left( \left | A^{(1,2)} \right | - 1 \right)}.$$
Optimal consensus clustering results of $G^{(1)}$ and $G^{(2)}$ will be $\hat{\mathcal{C}^{(1)}}, \hat{\mathcal{C}^{(2)}}$:
$$
\hat{\mathcal{C}}^{(1)}, \hat{\mathcal{C}}^{(2)} = \arg \min_{\mathcal{C}^{(1)}, \mathcal{C}^{(2)}} Nd(\mathcal{C}^{(1)}, \mathcal{C}^{(2)}).
$$





Similarly, the normalized-discrepancy objective function can also be represented with the \textit{clustering results confidence matrices} $\mb{H}^{(1)}$ and $\mb{H}^{(2)}$ as well. Meanwhile, considering that the networks studied in this paper are partially aligned, matrices $\mb{H}^{(1)}$ and $\mb{H}^{(2)}$ contain the results of both anchor users and non-anchor users, while non-anchor users should not be involved in the discrepancy calculation according to the definition of discrepancy. We propose to prune the results of the non-anchor users with the following \textit{anchor transition matrix} first.


\noindent \textbf{Definition 4} (Anchor Transition Matrix): Binary matrix $\mb{T}^{(1,2)}$ (or $\mb{T}^{(2,1)}$) is defined as the anchor transition matrix from networks $G^{(1)}$ to $G^{(2)}$ (or from $G^{(2)}$ to $G^{(1)}$), where $\mb{T}^{(1,2)} = ({\mb{T}^{(2,1)}})^T$, $\mb{T}^{(1,2)}(i, j) = 1$ if $(u^{(1)}_i, u^{(2)}_j) \in A^{(1,2)}$ and $0$ otherwise. The row indexes of $\mb{T}^{(1,2)}$ (or $\mb{T}^{(2,1)}$) are of the same order as those of $\mb{H}^{(1)}$ (or $\mb{H}^{(2)}$). Considering that the constraint on anchor links is ``\textit{one-to-one}'' in this paper, as a result, each row/column of $\mb{T}^{(1,2)}$ and $\mb{T}^{(2,1)}$ contains at most one entry filled with $1$.

%

\begin{figure}[t]
\centering
    \begin{minipage}[l]{1.0\columnwidth}
      \centering
      \includegraphics[width=1.0\textwidth]{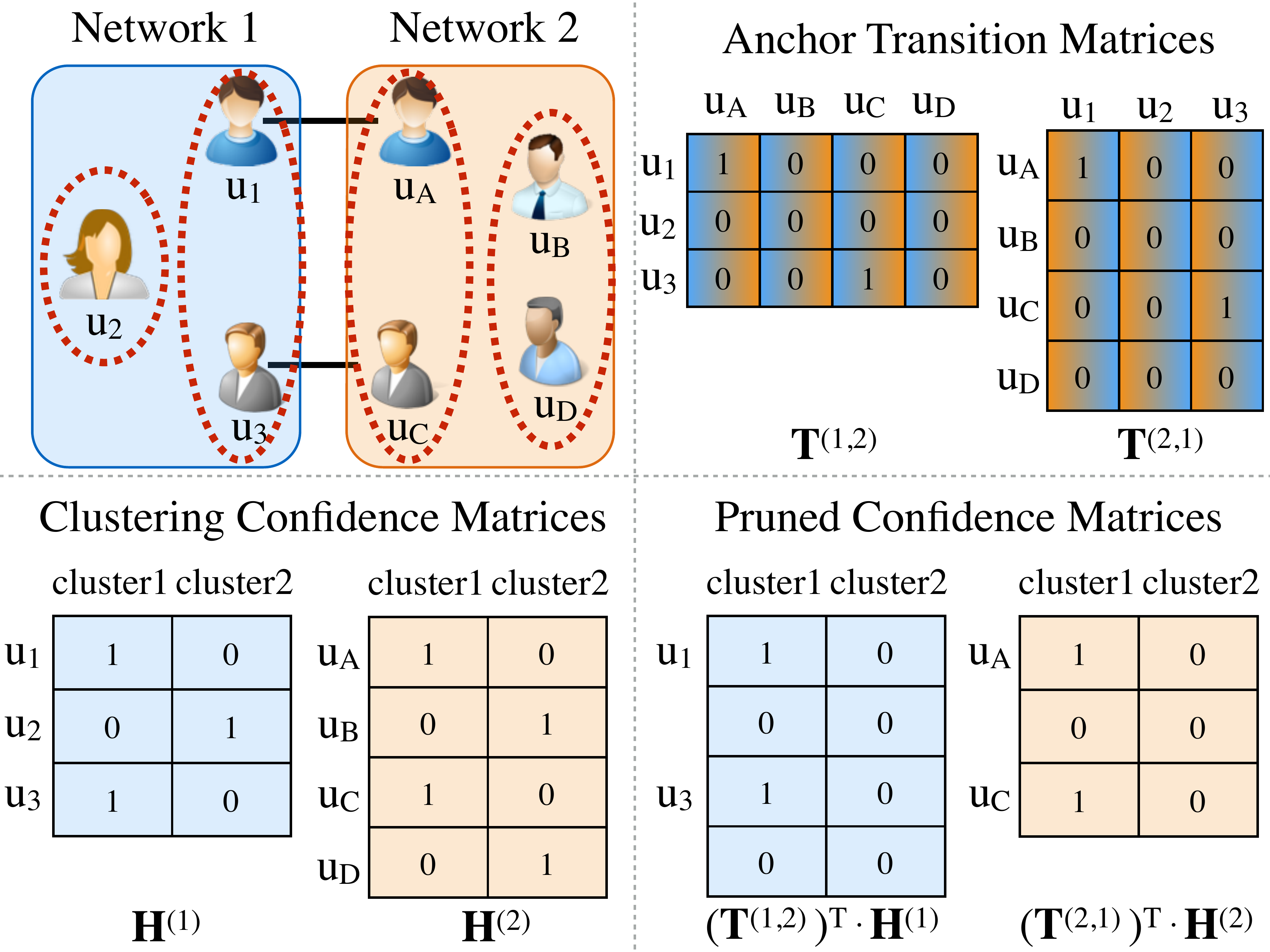}
    \end{minipage}
\caption{An example to illustrate the clustering discrepancy.}\label{fig_discrepancy_example}
\end{figure}

In Figure~\ref{fig_discrepancy_example}, we show an example about the clustering discrepancy of two partially aligned networks $G^{(1)}$ and $G^{(2)}$, users in which are grouped into two clusters $\{\{u_1, u_3\}, \{u_2\}\}$ and $\{\{u_A, u_C\}, \{u_B, u_D\}\}$ respectively. Users $u_1$, $u_A$ and $u_3$, $u_C$ are identified to be anchor users, based on which we can construct the ``anchor transition matrices'' $\mb{T}^{(1,2)}$ and $\mb{T}^{(2,1)}$ as shown in the upper right plot. Furthermore, based on the community structure, we can construct the ``\textit{clustering confidence matrices}'' as shown in the lower left plot. To obtain the clustering results of anchor users only, the \textit{anchor transition matrix} can be applied to prune the clustering results of non-anchor users from the \textit{clustering confidence matrices}. By multiplying the \textit{anchor transition matrices} $(\mb{T}^{(1,2)})^T$ and $(\mb{T}^{(2,1)})^T$ with \textit{clustering confidence matrices} $\mb{H}^{(1)}$ and $\mb{H}^{(2)}$ respectively, we can obtain the ``pruned confidence matrices'' as show in the lower right plot of Figure~\ref{fig_discrepancy_example}. Entries corresponding anchor users $u_1$, $u_3$, $u_A$ and $u_C$ are preserved but those corresponding to non-anchor users are all pruned. 

In this example, the clustering discrepancy of the partially aligned networks should be $0$ according to the above discrepancy definition. Meanwhile, networks $G^{(1)}$ and $G^{(2)}$ are of different sizes and the pruned confidence matrices are of different dimensions, e.g., $(\mb{T}^{(1,2)})^T\mb{H}^{(1)} \in \mathbb{R}^{4\times2}$ and $(\mb{T}^{(2,1)})^T\mb{H}^{(2)} \in \mathbb{R}^{3\times2}$. To represent the discrepancy with the clustering confidence matrices, we need to further accommodate the dimensions of different pruned \textit{clustering confidence matrices}. It can be achieved by multiplying one pruned \textit{clustering confidence matrices} with the corresponding \textit{anchor transition matrix} again, which will not prune entries but only adjust the matrix dimensions. Let $\bar{\mb{H}}^{(1)} =  (\mb{T}^{(1,2)})^T\mb{H}^{(1)}$ and $\bar{\mb{H}}^{(2)} = (\mb{T}^{(1,2)})^T (\mb{T}^{(2,1)})^T {\mb{H}^{(2)}}$. In the example, we can represent the clustering discrepancy to be
$$\left \| \bar{\mb{H}}^{(1)} \left(\bar{\mb{H}}^{(1)} \right )^T - \bar{\mb{H}}^{(2)} \left ( \bar{\mb{H}}^{(2)} \right)^T \right \|^2_F = 0,$$
where matrix $\bar{\mb{H}}\bar{\mb{H}}^T$ indicates whether pairs of anchor users are in the same cluster or not.


Furthermore, the objective function of inferring clustering confidence matrices, which can minimize the normalized discrepancy can be represented as follows
\begin{align*}
\min_{\mb{H}^{(1)}, \mb{H}^{(2)}} & \frac{\left \| \bar{\mb{H}}^{(1)} \left(\bar{\mb{H}}^{(1)} \right )^T - \bar{\mb{H}}^{(2)} \left ( \bar{\mb{H}}^{(2)} \right)^T \right \|^2_F}{\left \| \mb{T}^{(1,2)} \right \|^2_F \left (\left \| \mb{T}^{(1,2)} \right \|^2_F  - 1\right)}, \\
s.t. \ \ &(\mb{H}^{(1)})^T\mb{D}^{(1)}\mb{H}^{(1)} = \mb{I}, (\mb{H}^{(2)})^T\mb{D}^{(2)}\mb{H}^{(2)} = \mb{I}.
\end{align*}
where $\mb{D}^{(1)}$, $\mb{D}^{(2)}$ are the corresponding diagonal matrices of HNMP-Sim matrices of networks $G^{(1)}$ and $G^{(2)}$ respectively.



\begin{algorithm}[t]
\caption{Curvilinear Search Method ($\mathcal{CSM}$)}
\label{alg:update}
\begin{algorithmic}[1]
	\REQUIRE $\mb{X}_k$ $C_k$, $Q_k$ and function $\mathcal{F}$\\
\qquad	parameters $\mb{\epsilon} = \{\rho, \eta, \delta, \tau, \tau_m, \tau_M\}$\\
\ENSURE  $\mb{X}_{k + 1}$, $C_{k + 1}$, $Q_{k + 1}$\\

\STATE	$\mb{Y}(\tau) = \left( \mb{I} + \frac{\tau}{2}\mb{A}\right)^{-1} \left( \mb{I} - \frac{\tau}{2}\mb{A}\right) \mb{X}_k$
\WHILE{$\mathcal{F}\left( \mb{Y}(\tau)\right) \ge \mb{C}_k + \rho \tau \mathcal{F}^\prime \left( (\mb{Y}(0)) \right)$}
\STATE	{$\tau = \delta \tau$}
\STATE	$\mb{Y}(\tau) = \left( \mb{I} + \frac{\tau}{2}\mb{A}\right)^{-1} \left( \mb{I} - \frac{\tau}{2}\mb{A}\right) \mb{X}_k$
\ENDWHILE
\STATE	{$\mb{X}_{k + 1} = \mb{Y}_k(\tau)$\\$Q_{k + 1} = \eta Q_k + 1$\\ $C_{k + 1} = \left( \eta Q_k C_k + \mathcal{F}(\mb{X}_{k + 1}) \right)/Q_{k + 1}$\\$\tau = \max \left( \min(\tau, \tau_M), \tau_m \right)$}
\end{algorithmic}
\end{algorithm}

\subsection{Joint Mutual Clustering of Multiple Networks}\label{subsec:joint}
Normalized-Cut objective function favors clustering results that can preserve the characteristic of each network, however, normalized-discrepancy objective function favors consensus results which are mutually refined with information from other aligned networks. Taking both of these two issues into considerations, the optimal mutual clustering results $\mathcal{\hat{C}}^{(1)}$ and $\mathcal{\hat{C}}^{(2)}$ of aligned networks $G^{(1)}$ and $G^{(2)}$ can be achieved as follows:
\begin{align*}
\arg \min_{\mathcal{C}^{(1)}, \mathcal{C}^{(2)}} \alpha \cdot Ncut(\mathcal{C}^{(1)}) + \beta \cdot Ncut(\mathcal{C}^{(2)}) + \theta \cdot Nd(\mathcal{C}^{(1)}, \mathcal{C}^{(2)})
\end{align*}
where $\alpha$, $\beta$ and $\theta$ represents the weights of these terms and, for simplicity, $\alpha$, $\beta$ are both set as $1$ in this paper.

By replacing $Ncut(\mathcal{C}^{(1)})$, $Ncut(\mathcal{C}^{(2)})$, $Nd(\mathcal{C}^{(1)}, \mathcal{C}^{(2)})$ with the objective equations derived above, we can rewrite the joint objective function as follows:
\begin{align*}
\min_{\mb{H}^{(1)}, \mb{H}^{(2)}}\ \ \alpha \cdot &\mbox{Tr} (({\mb{H}^{(1)}})^T\mb{L}^{(1)}\mb{H}^{(1)}) + \beta \cdot \mbox{Tr} (({\mb{H}^{(2)}})^T\mb{L}^{(2)}\mb{H}^{(2)}) \\
&+ \theta \cdot \frac{\left \| \bar{\mb{H}}^{(1)} \left(\bar{\mb{H}}^{(1)} \right )^T - \bar{\mb{H}}^{(2)} \left ( \bar{\mb{H}}^{(2)} \right)^T \right \|^2_F}{\left \| \mb{T}^{(1,2)} \right \|^2_F \left (\left \| \mb{T}^{(1,2)} \right \|^2_F  - 1\right)},\\
s.t. \ \ &({\mb{H}^{(1)}})^T\mb{D}^{(1)}\mb{H}^{(1)} = \mb{I}, ({\mb{H}^{(2)}})^T\mb{D}^{(2)}\mb{H}^{(2)} = \mb{I},
\end{align*}
where $\mb{L}^{(1)} = \mb{D}^{(1)} - \mb{S}^{(1)}$, $\mb{L}^{(2)} = \mb{D}^{(2)} - \mb{S}^{(2)}$ and matrices $\mb{S}^{(1)}$, $\mb{S}^{(2)}$ and $\mb{D}^{(1)}$, $\mb{D}^{(2)}$ are the HNMP-Sim matrices and their corresponding diagonal matrices defined before.

The objective function is a complex optimization problem with orthogonality constraints, which can be very difficult to solve because the constraints are not only non-convex but also numerically expensive to preserve during iterations. Meanwhile, by substituting $\left( \mb{D}^{(1)} \right)^{\frac{1}{2}}\mb{H}^{(1)}$ and $\left( \mb{D}^{(2)} \right)^{\frac{1}{2}}\mb{H}^{(2)}$ with $\mb{X}^{(1)}$, $\mb{X}^{(2)}$, we can transform the objective function into a standard form of problems solvable with method proposed in \cite{WY10}:
\begin{align*}
&\min_{\mb{X}^{(1)}, \mb{X}^{(2)}}\ \ \alpha \cdot (\mbox{Tr} (({\mb{X}^{(1)}})^T \tilde{\mb{L}}^{(1)} \mb{X}^{(1)}) + \beta \cdot \mbox{Tr} (({\mb{X}^{(2)}})^T \tilde{\mb{L}}^{(2)} \mb{X}^{(2)}) \\
&+ \theta \cdot \frac{\left \| \tilde{\mb{T}}^{(1)}\mb{X}^{(1)} \left(\tilde{\mb{T}}^{(1)} \mb{X}^{(1)}\right )^T - \tilde{\mb{T}}^{(2)} \mb{X}^{(2)} \left (\tilde{\mb{T}}^{(2)} \mb{X}^{(2)} \right)^T \right \|^2_F}{\left \| \mb{T}^{(1,2)} \right \|^2_F \left (\left \| \mb{T}^{(1,2)} \right \|^2_F  - 1\right)}),\\
&s.t. \ \  ({\mb{X}^{(1)}})^T \mb{X}^{(1)} = \mb{I}, ({\mb{X}^{(2)}})^T \mb{X}^{(2)} = \mb{I}.
\end{align*}
where $\tilde{\mb{L}}^{(1)} = ( (\mb{D}^{(1)})^{-\frac{1}{2}} )^T \mb{L}^{(1)} ( (\mb{D}^{(1)})^{-\frac{1}{2}} )$,  $\tilde{\mb{L}}^{(2)} = ( (\mb{D}^{(2)})^{-\frac{1}{2}} )^T\\ \mb{L}^{(2)} ( (\mb{D}^{(2)})^{-\frac{1}{2}} )$ and $\tilde{\mb{T}}^{(1)} = (\mb{T}^{(1,2)})^T (\mb{D}^{(1)})^{-\frac{1}{2}}$,\\ $\tilde{\mb{T}}^{(2)} = (\mb{T}^{(1,2)})^T (\mb{T}^{(2,1)})^T (\mb{D}^{(2)})^{-\frac{1}{2}}$.

Wen et al. \cite{WY10} propose a feasible method to solve the above optimization problems with a constraint-preserving update scheme. They propose to update one variable, e.g., $\mb{X}^{(1)}$, while fixing the other variable, e.g., $\mb{X}^{(2)}$, alternatively with the curvilinear search with Barzilai-Borwein step method until convergence. For example, when $\mb{X}^{(2)}$ is fixed, we can simplify the objective function into
$$\min_{\mb{X}} \mathcal{F}(\mb{X}), s.t. (\mb{X})^T\mb{X} = \mb{I},$$
where $\mb{X} = \mb{X}^{(1)}$ and $\mathcal{F}(\mb{X})$ is the objective function, which can be solved with the curvilinear search with Barzilai-Borwein step method proposed in \cite{WY10} to update $\mb{X}$ until convergence and the variable $\mb{X}$ after the $(k+1)_{th}$ iteration will be
\begingroup\makeatletter\def\f@size{8}\check@mathfonts
$$\mb{X}_{k+1} = \mb{Y}(\tau_k), \mb{Y}(\tau_k) = \left( \mb{I} + \frac{\tau_k}{2}\mb{A}\right)^{-1} \left( \mb{I} - \frac{\tau_k}{2}\mb{A}\right) \mb{X}_k,$$
$$\mb{A} = \frac{\partial \mathcal{F}(\mb{X}_k)}{\partial \mb{X}} \mb{X}_k^T - \mb{X}_k (\frac{\partial \mathcal{F}(\mb{X}_k)}{\partial \mb{X}})^T,$$
\endgroup
where let $\hat{\tau} = \left(\frac{\mbox{Tr} \left( (\mb{X}_k - \mb{X}_{k - 1})^T  (\mb{X}_k - \mb{X}_{k - 1}) \right) }{ \left |\mbox{Tr} \left ( (\mb{X}_k - \mb{X}_{k - 1})^T \left (\nabla\mathcal{F}(\mb{X}_k) - \nabla\mathcal{F}(\mb{X}_{k - 1}) \right) \right ) \right |} \right)$, $\tau_k = \hat{\tau} \delta^h$, 
$\delta$ is the Barzilai-Borwein step size and $h$ is the smallest integer to make 
$\tau_k$ satisfy 
\begingroup\makeatletter\def\f@size{8}\check@mathfonts
$$\mathcal{F}\left( \mb{Y}(\tau_k) \right) \le C_k + \rho \tau_k \mathcal{F}^\prime_{\tau} \left( \mb{Y}(0) \right).$$
\endgroup
Terms $C$, $Q$ are defined as $C_{k + 1} = \left( \eta Q_k C_k + \mathcal{F}(\mb{X}_{k + 1}) \right) / Q_{k + 1}$ and $Q_{k + 1} = \eta Q_{k} + 1, Q_0 = 1$. More detailed derivatives of the curvilinear search method (i.e., Algorithm~\ref{alg:update}) with Barzilai-Borwein step is available in \cite{WY10}. Meanwhile, the pseudo-code of method {\our} is available in Algorithm~\ref{alg:framework}. Based on the achieved solutions $\mb{X}^{(1)}$ and $\mb{X}^{(2)}$, we can get $\mb{H}^{(1)} = \left( \mb{D}^{(1)} \right)^{-\frac{1}{2}} \mb{X}^{(1)}$ and $\mb{H}^{(2)} = \left( \mb{D}^{(2)} \right)^{-\frac{1}{2}} \mb{X}^{(2)}$.


\begin{algorithm}[t]
\caption{Mutual Community Detector ({\our})}
\label{alg:framework}
\begin{algorithmic}[1]
	\REQUIRE aligned network: $\mathcal{G}$ = $\{\{G^{(1)}$, $G^{(2)}\}$, $\{A^{(1,2)},$ $A^{(2,1)}\}\}$;\\
\qquad	number of clusters in $G^{(1)}$ and $G^{(2)}$: $k^{(1)}$ and $k^{(2)}$;\\
\qquad	HNMP Sim matrices weight: $\mb{\omega}$;\\
\qquad	parameters: $\mb{\epsilon} = \{\rho, \eta, \delta, \tau, \tau_m, \tau_M\}$;\\
\qquad	function $\mathcal{F}$ and consensus term weight $\theta$
\ENSURE  $\mb{H}^{(1)}$, $\mb{H}^{(2)}$\\
\STATE	Calculate HNMP Sim matrices, $\mb{S}_i^{(1)}$ and $\mb{S}^{(2)}_i$
\STATE	$\mb{S}^{(1)} = \sum_i \omega_i S_i^{(1)}$, $\mb{S}^{(2)} = \sum_i \omega_i S_i^{(2)}$
\STATE	Initialize $\mb{X}^{(1)}$ and $\mb{X}^{(2)}$ with Kmeans clustering results on $\mb{S}^{(1)}$ and $\mb{S}^{(2)}$
\STATE	Initialize $C^{(1)}_0 = 0, Q^{(1)}_0 = 1$ and $C^{(2)}_0 = 0, Q^{(2)}_0 = 1$
\STATE	$converge = False$
\WHILE{$converge = False$}
\STATE	/* update $\mb{X}^{(1)}$ and $\mb{X}^{(2)}$ with $\mathcal{CSM}$ */\\
	$\mb{X}^{(1)}_{k+1}$, $C^{(1)}_{k+1}$, $Q^{(1)}_{k+1}$ = $\mathcal{CSM}(\mb{X}^{(1)}_k, C^{(1)}_k, Q^{(1)}_k, \mathcal{F}, \mb{\epsilon})$\\
		$\mb{X}^{(2)}_{k+1}$, $C^{(2)}_{k+1}$, $Q^{(2)}_{k+1}$ = $\mathcal{CSM}(\mb{X}^{(2)}_k, C^{(2)}_k, Q^{(2)}_k, \mathcal{F}, \mb{\epsilon})$
\IF{$\mb{X}^{(1)}_{k+1}$ and $\mb{X}^{(2)}_{k+1}$ both converge}
\STATE	$converge = True$ 
\ENDIF
\ENDWHILE
\STATE	{$\mb{H}^{(1)} = \left((\mb{D}^{(1)})^{-\frac{1}{2}} \right)^T \mb{X}^{(1)}$, $\mb{H}^{(2)} = \left((\mb{D}^{(2)})^{-\frac{1}{2}} \right)^T \mb{X}^{(2)}$}
\end{algorithmic}
\end{algorithm}



\section{Experiments}\label{sec:experiment}
To demonstrate the effectiveness of {\our}, we will conduct extensive experiments on two real-world partially aligned heterogeneous networks: Foursquare and Twitter, in this section.

\subsection{Dataset Description}

\begin{table}[t]
\caption{Properties of the Heterogeneous Social Networks}
\label{tab:datastat}
\centering
\footnotesize
\begin{tabular}{clrr}
\toprule
&&\multicolumn{2}{c}{network}\\
\cmidrule{3-4}
&property &\textbf{Twitter}	&\textbf{Foursquare}	 \\
\midrule 
\multirow{3}{*}{\# node}
&user		& 5,223	& 5,392 \\
&tweet/tip	& 9,490,707	& 48,756 \\
&location	& 297,182	& 38,921 \\
\midrule 
\multirow{3}{*}{\# link}
&friend/follow		&164,920	&76,972 \\
&write		& 9,490,707	& 48,756 \\
&locate		& 615,515	& 48,756 \\
\bottomrule
\end{tabular}
\end{table}

As mentioned in the Section~\ref{sec:formulation}, both Foursquare and Twitter used in this paper are heterogeneous social networks, whose statistical information is given in Table~\ref{tab:datastat}. These two networks were crawled with the methods proposed in \cite{KZY13} during November, 2012. The number of anchor links obtained is $3,388$. Some basic descriptions about datasets are as follows:

\begin{itemize}
\item \textbf{Foursquare}: Users together with their posts are crawled from Foursquare, whose number are $5,392$ and $48,756$ respectively. The number of social link among users is $76,972$. All these posts written by these users and can attach locations checkins and, as a result, the numbers of write link and locate link are both $48,756$. $38,921$ different locations are crawled from Foursquare.

\item \textbf{Twitter}: $5,223$ users and all their tweets, whose number is $9,490,707$, are crawled from Twitter and, on average, each user has about $1,817$ tweets. Among these tweets, about $615,515$ have location check-ins, which accounts for about $6.48\%$ of all tweets. The number of locations crawled from Twitter is $297,182$ and the number of social links among users is $164,920$.

\end{itemize}

For more information about the datasets and crawling methods, please refer to \cite{KZY13, ZKY13, ZKY14, ZYZ14}.



\begin{table*}[t]

\caption{Community Detection Results of Foursquare and Twitter Evaluated by Quality Metrics.}
\label{tab:setting1}
\centering
{
\scriptsize
\begin{tabular}{llrcccccccccc}
\toprule
\multicolumn{2}{l}{ }&\multicolumn{10}{c}{remaining anchor link rates $\sigma$}\\
\cmidrule{4-13}
network &measure &methods &0.1	& 0.2	& 0.3	& 0.4	&0.5	&0.6	&0.7	& 0.8	 &0.9	&1.0\\
\midrule
\multirow{16}{*}{\rotatebox{90}{Foursquare}}

&\multirow{4}{*}{ndbi}
&{\our}     &\textbf{0.927}     &\textbf{0.924}     &\textbf{0.95}     &\textbf{0.969}     &\textbf{0.966}     &\textbf{0.961}     &\textbf{0.958}     &\textbf{0.954}     &\textbf{0.971}     &\textbf{0.958}     \\
&&{\ourinfluence}     &0.891     &0.889     &0.88     &0.877     &0.894     &0.883     &0.89     &0.88     &0.887     &0.893      \\
&&{\ourcut}     &0.863     &0.863     &0.863     &0.863     &0.863     &0.863     &0.863     &0.863     &0.863     &0.863     \\
&&{\ourkmeans}     &0.835     &0.835     &0.835     &0.835     &0.835     &0.835     &0.835     &0.835     &0.835     &0.835     \\

\cmidrule{2-13}

&\multirow{4}{*}{entropy}
&{\our}     &\textbf{1.551}     &\textbf{1.607}     &\textbf{1.379}     &\textbf{1.382}     &\textbf{1.396}     &\textbf{1.382}     &\textbf{1.283}     &\textbf{1.552}     &\textbf{1.308}     &\textbf{1.497}     \\
&&{\ourinfluence}     &4.332     &4.356     &4.798     &4.339     &4.474     &4.799     &4.446     &4.658     &4.335     &4.459     \\
&&{\ourcut}     &2.768     &2.768     &2.768     &2.768     &2.768     &2.768     &2.768     &2.768     &2.768     &2.768     \\
&&{\ourkmeans}     &2.369     &2.369     &2.369     &2.369     &2.369     &2.369     &2.369     &2.369     &2.369     &2.369     \\

\cmidrule{2-13}

&\multirow{4}{*}{density}
&{\our}     &\textbf{0.216}     &\textbf{0.205}     &\textbf{0.196}     &0.163    &\textbf{0.239}     &\textbf{0.192}     &\textbf{0.303}     &\textbf{0.198}     &0.170    &\textbf{0.311}     \\
&&{\ourinfluence}     &0.116     &0.121     &0.13     &0.095     &0.143     &0.11     &0.13     &0.12     &0.143     &0.103     \\
&&{\ourcut}     &0.154     &0.154     &0.154     &0.154     &0.154     &0.154     &0.154     &0.154     &0.154     &0.154     \\
&&{\ourkmeans}     &0.182     &0.182     &0.182     &\textbf{0.182}     &0.182     &0.182     &0.182     &0.182     &\textbf{0.182}     &0.182     \\

\cmidrule{2-13}

&\multirow{4}{*}{silhouette}
&{\our}     &\textbf{-0.137}     &\textbf{-0.114}     &\textbf{-0.148}     &\textbf{-0.156}     &\textbf{-0.117}     &\textbf{-0.11}     &\textbf{-0.035}     &\textbf{-0.125}     &\textbf{-0.148}     &\textbf{-0.044}     \\
&&{\ourinfluence}     &-0.168     &-0.198     &-0.173     &-0.189     &-0.178     &-0.181     &-0.21     &-0.195     &-0.167     &-0.18     \\
&&{\ourcut}     &-0.34     &-0.34     &-0.34     &-0.34     &-0.34     &-0.34     &-0.34     &-0.34     &-0.34     &-0.34     \\
&&{\ourkmeans}     &-0.297     &-0.297     &-0.297     &-0.297     &-0.297     &-0.297     &-0.297     &-0.297     &-0.297     &-0.297     \\

\midrule

\multirow{16}{*}{\rotatebox{90}{Twitter}}
&\multirow{4}{*}{ndbi}
&{\our}     &\textbf{0.962}     &\textbf{0.969}     &\textbf{0.955}     &\textbf{0.969}     &\textbf{0.97}     &\textbf{0.958}     &\textbf{0.952}     &\textbf{0.96}     &\textbf{0.946}     &\textbf{0.953}     \\
&&{\ourinfluence}     &0.815     &0.843     &0.807     &0.83     &0.826     &0.832     &0.835     &0.808     &0.812     &0.836     \\
&&{\ourcut}     &0.759     &0.759     &0.759     &0.759     &0.759     &0.759     &0.759     &0.759     &0.759     &0.759     \\
&&{\ourkmeans}     &0.761     &0.761     &0.761     &0.761     &0.761     &0.761     &0.761     &0.761     &0.761     &0.761     \\

\cmidrule{2-13}

&\multirow{4}{*}{entropy}
&{\our}     &\textbf{2.27}     &\textbf{2.667}     &\textbf{2.48}     &\textbf{2.381}     &\textbf{2.43}     &\textbf{2.372}     &\textbf{2.452}     &\textbf{2.459}     &\textbf{2.564}     &\textbf{2.191}     \\
&&{\ourinfluence}     &4.780     &5.114     &5.066     &4.961     &4.904     &4.866     &5.121     &4.629     &4.872     &5.000     \\
&&{\ourcut}     &3.099     &3.099     &3.099     &3.099     &3.099     &3.099     &3.099     &3.099     &3.099     &3.099     \\
&&{\ourkmeans}     &3.245     &3.245     &3.245     &3.245     &3.245     &3.245     &3.245     &3.245     &3.245     &3.245     \\

\cmidrule{2-13}

&\multirow{4}{*}{density}
&{\our}     &\textbf{0.14}     &0.097     &\textbf{0.142}     &0.109     &\textbf{0.15}     &\textbf{0.158}     &\textbf{0.126}     &\textbf{0.149}     &\textbf{0.147}     &\textbf{0.164}     \\
&&{\ourinfluence}     &0.055     &0.017     &0.044     &0.026     &0.04     &0.062     &0.016     &0.044     &0.045     &0.02     \\
&&{\ourcut}     &0.107     &0.107     &0.107     &0.107     &0.107     &0.107     &0.107     &0.107     &0.107     &0.107     \\
&&{\ourkmeans}     &0.119     &\textbf{0.119}     &0.119     &\textbf{0.119}     &0.119     &0.119     &0.119     &0.119     &0.119     &0.119     \\

\cmidrule{2-13}

&\multirow{4}{*}{silhouette}
&{\our}     &\textbf{-0.137}     &\textbf{-0.179}     &\textbf{-0.282}     &\textbf{-0.175}     &\textbf{-0.275}     &\textbf{-0.273}     &\textbf{-0.248}     &\textbf{-0.269}     &\textbf{-0.266}     &\textbf{-0.286}     \\
&&{\ourinfluence}     &-0.356     &-0.322     &-0.311     &-0.347     &-0.346     &-0.349     &-0.323     &-0.363     &-0.345     &-0.352     \\
&&{\ourcut}     &-0.424     &-0.424     &-0.424     &-0.424     &-0.424     &-0.424     &-0.424     &-0.424     &-0.424     &-0.424     \\
&&{\ourkmeans}     &-0.406     &-0.406     &-0.406     &-0.406     &-0.406     &-0.406     &-0.406     &-0.406     &-0.406     &-0.406     \\

\bottomrule

\end{tabular}
}
\end{table*}
\begin{table*}[t]
\caption{Community Detection Results of Foursquare and Twitter Evaluated by Consensus Metrics.}
\label{tab:setting3}
\centering
{\scriptsize
\begin{tabular}{lrcccccccccc}
\toprule
\multicolumn{2}{l}{ }&\multicolumn{10}{c}{remaining anchor link rates $\sigma$}\\
\cmidrule{3-12}
measure &methods  &0.1	& 0.2	& 0.3	& 0.4	&0.5	&0.6	&0.7	& 0.8	 &0.9	&1.0\\
\midrule

\multirow{4}{*}{rand}
&{\our}	&\textbf{0.095}     &\textbf{0.099}     &\textbf{0.107}     &\textbf{0.138}     &\textbf{0.116}     &\textbf{0.121}     &\textbf{0.132}     &\textbf{0.106}     &\textbf{0.089}     &0.159\\

&{\ourinfluence}	&0.135     &0.139     &0.144     &0.148     &0.142     &0.14     &\textbf{0.132}     &0.132     &0.144     &\textbf{0.141} \\

&{\ourcut} &0.399     &0.377     &0.372     &0.4     &0.416     &0.423     &0.362     &0.385     &0.362     &0.341   \\

&{\ourkmeans}	&0.436     &0.387     &0.4     &0.358     &0.403     &0.363     &0.408     &0.365     &0.35     &0.363 \\

\midrule   

\multirow{4}{*}{vi}
&{\our}	&\textbf{3.309}     &\textbf{4.052}     &\textbf{4.058}     &\textbf{3.902}     &\textbf{4.038}     &\textbf{4.348}     &\textbf{3.973}     &\textbf{3.944}     &\textbf{4.078}     &\textbf{2.911}  \\

&{\ourinfluence}	&7.56     &8.324     &8.414     &8.713     &8.756     &8.836     &8.832     &8.621     &8.427     &8.02\\

&{\ourcut} &5.384     &5.268     &5.221     &4.855     &5.145     &5.541     &5.909     &5.32     &5.085     &5.246\\

&{\ourkmeans}	&5.427     &5.117     &5.355     &5.326     &5.679     &5.944     &5.452     &5.567     &5.513     &4.686\\

\midrule

\multirow{4}{*}{nmi}
&{\our}	&0.152     &\textbf{0.152}     &\textbf{0.149}     &\textbf{0.141}     &\textbf{0.149}     &\textbf{0.156}     &\textbf{0.142}     &\textbf{0.158}     &\textbf{0.147}     &0.146\\

&{\ourinfluence}	&\textbf{0.172}     &0.097     &0.081     &0.06     &0.056     &0.069     &0.078     &0.093     &0.105     &\textbf{0.149}\\

&{\ourcut} &0.075     &0.074     &0.111     &0.108     &0.109     &0.099     &0.05     &0.036     &0.042     &0.106\\

&{\ourkmeans}	&0.008     &0.047     &0.048     &0.054     &0.048     &0.028     &0.047     &0.014     &0.067     &0.119\\

\midrule

\multirow{4}{*}{mi}
&{\our}	&0.756     &\textbf{0.611}     &\textbf{0.4}     &0.258     &\textbf{0.394}     &\textbf{0.431}     &\textbf{0.381}     &\textbf{0.533}     &\textbf{0.697}     &0.689\\

&{\ourinfluence}	&\textbf{0.780}     &0.446     &0.367     &\textbf{0.277}     &0.258     &0.325     &0.374     &0.44     &0.489     &\textbf{0.698} \\

&{\ourcut} &0.188     &0.181     &0.261     &0.232     &0.252     &0.243     &0.138     &0.092     &0.111     &0.31\\

&{\ourkmeans}	&0.02     &0.112     &0.119     &0.135     &0.127     &0.078     &0.119     &0.038     &0.194     &0.314\\

\bottomrule

\end{tabular}
}
\end{table*}

\begin{table*}[t]
\caption{Community Detection Results of Foursquare and Twitter Evaluated by $IQC$ Metrics.}
\label{tab:setting4}
\centering
{\scriptsize
\begin{tabular}{lrcccccccccc}
\toprule
\multicolumn{2}{l}{ }&\multicolumn{10}{c}{remaining anchor link rates $\sigma$}\\
\cmidrule{2-12}
measure &methods &0.1	& 0.2	& 0.3	& 0.4	&0.5	&0.6	&0.7	& 0.8	 &0.9	&1.0\\
\midrule

\multirow{4}{*}{$IQC^{ndbi}_{rand}$}
&{\our}     &\textbf{-1.699}     &\textbf{-1.695}     &\textbf{-1.691}     &\textbf{-1.662}     &\textbf{-1.705}     &\textbf{-1.676}     &\textbf{-1.647}     &\textbf{-1.703}     &\textbf{-1.738}     &\textbf{-1.594}     \\
&{\ourinfluence}     &-1.459     &-1.451     &-1.44     &-1.434     &-1.444     &-1.45     &-1.465     &-1.465     &-1.442     &-1.448     \\
&{\ourcut}     &-0.824     &-0.869     &-0.878     &-0.821     &-0.789     &-0.776     &-0.899     &-0.851     &-0.897     &-0.94     \\
&{\ourkmeans}     &-0.724     &-0.821     &-0.795     &-0.88     &-0.79     &-0.87     &-0.779     &-0.865     &-0.895     &-0.869     \\





\cmidrule{2-12}

\multirow{4}{*}{$IQC^{ent.}_{vi}$}
&{\our}     &\textbf{10.439}     &\textbf{12.379}     &\textbf{11.975}     &\textbf{11.566}     &\textbf{11.902}     &\textbf{12.45}     &\textbf{11.681}     &\textbf{11.897}     &\textbf{12.028}     &\textbf{9.509}     \\
&{\ourinfluence}     &24.58     &26.107     &26.287     &26.884     &26.971     &27.13     &27.123     &26.7     &26.313     &25.499     \\
&{\ourcut}     &16.634     &16.403     &16.308     &15.577     &16.156     &16.948     &17.684     &16.506     &16.036     &16.359     \\
&{\ourkmeans}     &16.468     &15.847     &16.325     &16.267     &16.972     &17.503     &16.519     &16.748     &16.641     &14.986     \\

\cmidrule{2-12}

\multirow{4}{*}{$IQC^{dens.}_{nmi}$}
&{\our}     &\textbf{-0.659}     &\textbf{-0.606}     &\textbf{-0.636}     &\textbf{-0.555}     &\textbf{-0.686}     &\textbf{-0.663}     &\textbf{-0.713}     &\textbf{-0.664}     &\textbf{-0.611}     &\textbf{-0.768}     \\
&{\ourinfluence}     &-0.467     &-0.317     &-0.284     &-0.243     &-0.235     &-0.261     &-0.28     &-0.309     &-0.332     &-0.421     \\
&{\ourcut}     &-0.411     &-0.409     &-0.484     &-0.477     &-0.478     &-0.458     &-0.361     &-0.333     &-0.345     &-0.473     \\
&{\ourkmeans}     &-0.317     &-0.395     &-0.397     &-0.41     &-0.398     &-0.357     &-0.396     &-0.329     &-0.436     &-0.54     \\

\cmidrule{2-12}

\multirow{4}{*}{$IQC^{sil.}_{mi}$}
&{\our}     &\textbf{-1.239}     &\textbf{-0.93}     &\textbf{-0.371}     &\textbf{-0.186}     &\textbf{-0.396}     &\textbf{-0.479}     &\textbf{-0.479}     &\textbf{-0.673}     &\textbf{-0.979}     &\textbf{-1.048}     \\
&{\ourinfluence}     &-1.028     &-0.361     &-0.202     &-0.022     &0.016     &-0.118     &-0.216     &-0.347     &-0.446     &-0.863     \\
&{\ourcut}     &0.389     &0.403     &0.242     &0.3     &0.261     &0.278     &0.488     &0.58     &0.542     &0.144     \\
&{\ourkmeans}     &0.664     &0.479     &0.465     &0.433     &0.45     &0.546     &0.466     &0.628     &0.316     &0.074     \\

\bottomrule

\end{tabular}
}
\end{table*}


\subsection{Experiment Settings}
\subsubsection{Comparison Methods}
The comparison methods used in the experiments can be divided into three categories,

\noindent \textbf{Mutual Clustering Methods}
\begin{itemize}
\item {\our}: {\our} is the mutual community detection method proposed in this paper, which can detect the communities of multiple aligned networks with consideration of the connections and characteristics of different networks. Heterogeneous information in multiple aligned networks are applied in building {\our}.
\end{itemize}

\noindent \textbf{Multi-Network Clustering Methods}
\begin{itemize}
\item {\ourinfluence}: the clustering method proposed in \cite{ZL13, ZY15} can calculate the similarity scores among users by propagating heterogeneous information across views/networks. In this paper, we extend the method proposed in \cite{ZL13, ZY15} and propose {\ourinfluence} to calculate the intimacy scores among users in multiple networks simultaneously, based on which, users can be grouped into different clusters with clustering models based on intimacy matrix factorization as introduced in \cite{ZY15}. Heterogeneous information across networks is used to build {\ourinfluence}.
\end{itemize}

\noindent \textbf{Isolated Clustering Methods}, which can detect communities in each isolated network:
\begin{itemize}
\item {\ourcut}: {\ourcut} is the clustering method based on normalized cut proposed in \cite{SM00}. Method {\ourcut} can detect the communities in each social network merely based on the social connections in each network in the experiments.

\item {\ourkmeans}: {\ourkmeans} is a traditional clustering method, which can be used to detect communities \cite{QAH12} in social networks based on the social connections only in the experiments.
\end{itemize}

\subsubsection{Evaluation Methods}

The evaluation metrics applied in this paper can be divided into two categories: Quality Metrics and Consensus Metrics.

\noindent \textbf{Quality Metrics}: $4$ widely and commonly used quality metrics are applied to measure the clustering result, e.g., $\mathcal{C} = \{U_i\}_{i=1}^K$, of each network.

\begin{itemize}
\item \textit{normalized-dbi} \cite{ZL13}: 
$$ndbi(\mathcal{C}) = \frac{1}{K} \sum_{i} \min_{j \ne i} \\ \frac{d(c_i, c_j) + d(c_j, c_i)}{\sigma_i + \sigma_j + d(c_i, c_j) + d(c_j, c_i)},$$ 
where $c_i$ is the centroid of community $U_i \in \mathcal{C}$, $d(c_i, c_j)$ denotes the distance between centroids $c_i$ and $c_j$ and $\sigma_i$ represents the average distance between elements in $U_i$ and centroid $c_i$. (Higher ndbi corresponds to better performance).

\item \textit{entropy} \cite{ZL13}: 
$$H(\mathcal{C}) = - \sum_{i = 1}^K P(i) \log P(i),$$ where $P(i) = \frac{|U_i|}{\sum_{i = 1}^K |U_i|}$. (Lower entropy corresponds to better performance).

\item \textit{density} \cite{ZL13}: 
$$dens(\mathcal{C}) = \sum_{i = 1}^K \frac{|E_i|}{|E|},$$ where $E$ and $E_i$ are the edge sets in the network and $U_i$. (Higher density corresponds to better performance).

\item \textit{silhouette} \cite{LLXGW10}: 
$$sil(\mathcal{C}) = \frac{1}{K} \sum_{i = 1}^K (\frac{1}{|U_i|} \sum_{u \in U_i} \frac{b(u) - a(u)}{\max\{a(u), b(u)\}}),$$
where $a(u) = \frac{1}{|U_i| - 1}\sum_{v \in U_i, u \ne v} d(u,v)$ and \\$b(u) = \min_{j, j \ne i}\left( \frac{1}{|U_j|}\sum_{v \in U_j}d(u,v) \right)$. (Higher silhouette corresponds to better performance).
\end{itemize}

\noindent \textbf{Consensus Metrics}: Given the clustering results $\mathcal{C}^{(1)} = \{U^{(1)}_i\}_{i = 1}^{K^{(1)}}$ and $\mathcal{C}^{(2)} = \{U^{(2)}_i\}_{i = 1}^{K^{(2)}}$, the consensus metrics measuring the how similar or dissimilar the anchor users are clustered in $\mathcal{C}^{(1)}$ and $\mathcal{C}^{(2)}$ include:

\begin{itemize}
\item \textit{rand} \cite{NC07}: $rand(\mathcal{C}^{(1)}, \mathcal{C}^{(2)}) = \frac{N_{01} + N_{10}}{N_{00} + N_{01} + N_{10} + N_{11}}$, where $N_{11}$($N_{00}$) is the numbers of pairwise anchor users who are clustered in the same (different) community(ies) in both $\mathcal{C}^{(1)}$ and $\mathcal{C}^{(2)}$, $N_{01}$($N_{10}$) is that of anchor users who are clustered in the same community (different communities) in $\mathcal{C}^{(1)}$ but in different communities (the same communities) in $\mathcal{C}^{(2)}$. (Lower rand corresponds to better performance).

\item \textit{variation of information} \cite{NC07}: 
$$vi(\mathcal{C}^{(1)}, \mathcal{C}^{(2)}) = H(\mathcal{C}^{(1)}) + H(\mathcal{C}^{(2)}) - 2mi(\mathcal{C}^{(1)}, \mathcal{C}^{(2)}).$$ (Lower vi corresponds to better performance).

\item \textit{mutual information} \cite{NC07}: 
$$mi(\mathcal{C}^{(1)}, \mathcal{C}^{(2)}) = \sum_{i = 1}^{K^{(1)}} \sum_{j = 1}^{K^{(2)}} P(i,j) \log \frac{P(i,j)}{P(i)P(j)},$$ where $P(i,j) = \frac{|U_i^{(1)} \cap_{\mathcal{A}} U^{(2)}_j|}{|\mathcal{A}|}$ and $|U_i^{(1)} \cap_{\mathcal{A}} U^{(2)}_j| = \left| \{u | u \in U_i^{(1)}, \exists v \in U_i^{(2)}, (u, v) \in \mathcal{A}\} \right|$ \cite{KZY13}. (Higher mi corresponds to better performance).

\item \textit{normalized mutual information} \cite{NC07}: 
$$nmi(\mathcal{C}^{(1)}, \mathcal{C}^{(2)}) = \frac{mi(\mathcal{C}^{(1)}, \mathcal{C}^{(2)})}{\sqrt{H(\mathcal{C}^{(1)}) H(\mathcal{C}^{(2)})}}.$$ (Higher nmi corresponds to better performance).
\end{itemize}

\noindent \textbf{Definition 9} (Proportional and Inversely Proportional Metrics): Depending on relationship between the metric value and the clustering results, all the above metrics can be either \textit{proportional} or \textit{inversely proportional}. Metric $M$ is proportional iff better clustering results corresponds to higher $M$ value; $M$ is inversely proportional iff better clustering result corresponds lower $M$ value. 

In the metrics introduced above, \textit{normalized-dbi}, \textit{density}, \textit{silhouette}, \textit{mutual information} and \textit{normalized mutual information} are \textit{proportional metrics}, \textit{entropy}, \textit{rand}, and \textit{variation of information} are \textit{inversely proportional metrics}.

To consider both the quality and consensus simultaneously, we introduce a new clustering metric, $IQC$ metrics, in this paper, which is \textit{inversely proportional}.

\noindent \textbf{Definition 10} ($IQC$ metrics): $IQC$ is a linear combination of quality metrics $Q$ and consensus metrics $C$.
\begin{align*}
IQC(\mathcal{C}^{(1)}, \mathcal{C}^{(2)}) &= I(Q) (\beta_1 Q(\mathcal{C}^{(1)}) + \beta_2 Q(\mathcal{C}^{(2)}))\\
&+ I(C)(\beta_3 C(\mathcal{C}^{(1)}, \mathcal{C}^{(2)}) + \beta_4 C(\mathcal{C}^{(2)}, \mathcal{C}^{(1)}))
\end{align*}
where $\beta_1, \beta_2, \beta_3, \beta_4$ are weights of different terms, which are all set as $1$ in this paper, and $I(Q), I(C) = -1$, if $Q/C$ is proportional and $1$, otherwise.


\noindent \textbf{IQC Metrics} used in this paper include:

\begin{itemize}

\item $IQC^{ent}_{vi}(\mathcal{C}^{(1)}, \mathcal{C}^{(2)}) = H(\mathcal{C}^{(1)}) + H(\mathcal{C}^{(2)}) + 2 vi(\mathcal{C}^{(1)}, \mathcal{C}^{(2)})$

\item $IQC^{sil}_{mi}(\mathcal{C}^{(1)}, \mathcal{C}^{(2)}) = -sil(\mathcal{C}^{(1)}) - sil(\mathcal{C}^{(2)}) - 2 mi(\mathcal{C}^{(1)}, \mathcal{C}^{(2)})$

\item $IQC^{ndbi}_{rand}(\mathcal{C}^{(1)}, \mathcal{C}^{(2)}) = - ndbi(\mathcal{C}^{(1)}) - ndbi(\mathcal{C}^{(2)}) \\+ 2 rand(\mathcal{C}^{(1)}, \mathcal{C}^{(2)})$

\item $IQC^{dens}_{nmi}(\mathcal{C}^{(1)}, \mathcal{C}^{(2)}) = - dens(\mathcal{C}^{(1)}) - dens(\mathcal{C}^{(2)}) \\- 2 nmi(\mathcal{C}^{(1)}, \mathcal{C}^{(2)})$

\end{itemize}


\begin{figure}[t]
\centering
\subfigure[$\left \| \mb{X}^{(1)} \right \|_1$]{\label{eg_fig8_1}
    \begin{minipage}[l]{0.45\columnwidth}
      \centering
      \includegraphics[width=1.0\textwidth]{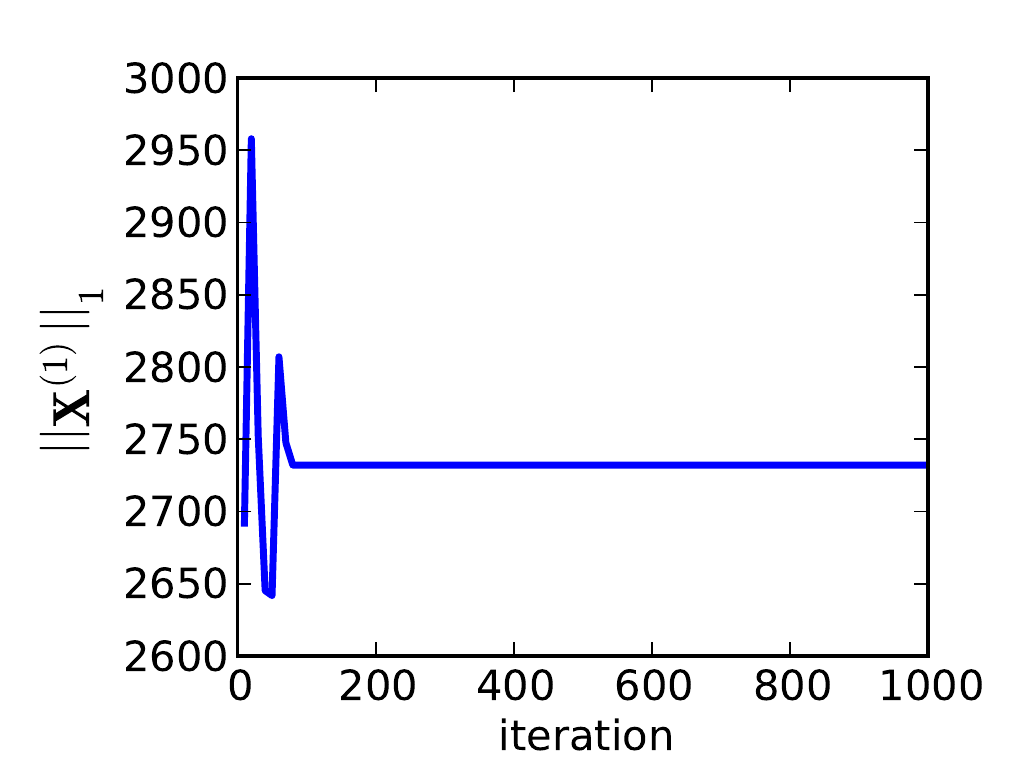}
    \end{minipage}
}
\subfigure[$\left \|\mb{X}^{(2)} \right \|_1$]{ \label{eg_fig8_2}
    \begin{minipage}[l]{0.45\columnwidth}
      \centering
      \includegraphics[width=1.0\textwidth]{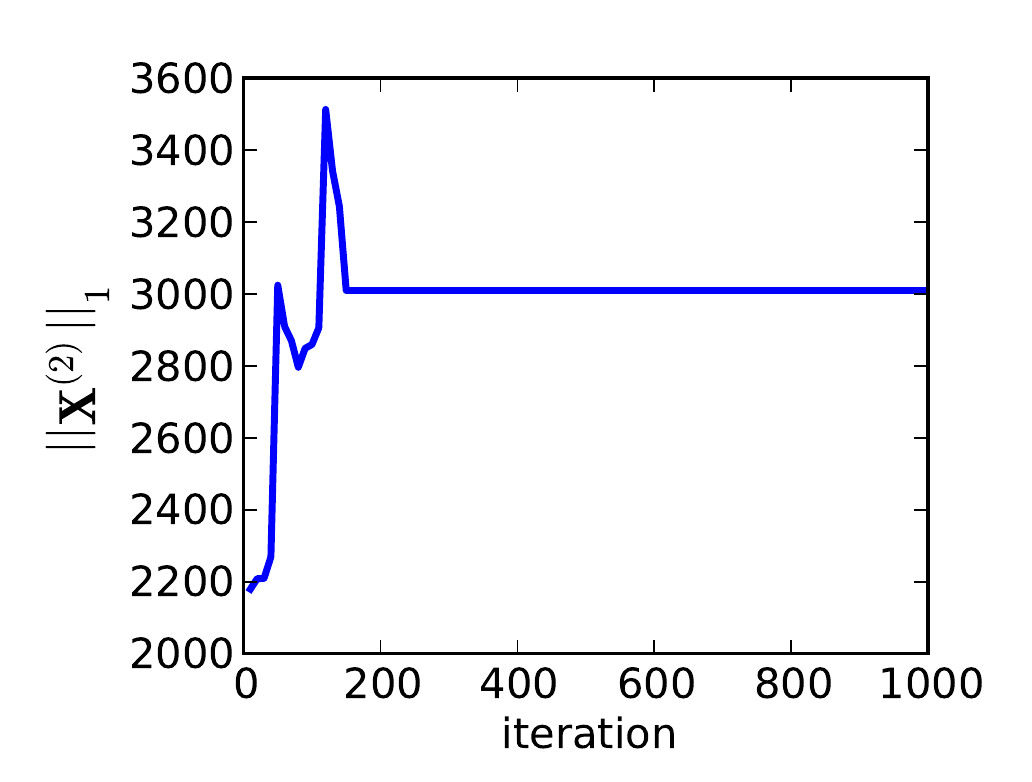}
    \end{minipage}
}
\caption{$\left \|\mb{X}^{(1)} \right \|_1$ and $\left \|\mb{X}^{(2)} \right \|_1$ in each iteration.}\label{eg_fig8}
\end{figure}

\begin{figure}[t]
\centering
\subfigure[$k^{(1)}$-ndbi (Foursquare)]{\label{eg_fig9_1}
    \begin{minipage}[l]{0.45\columnwidth}
      \centering
      \includegraphics[width=1.0\textwidth]{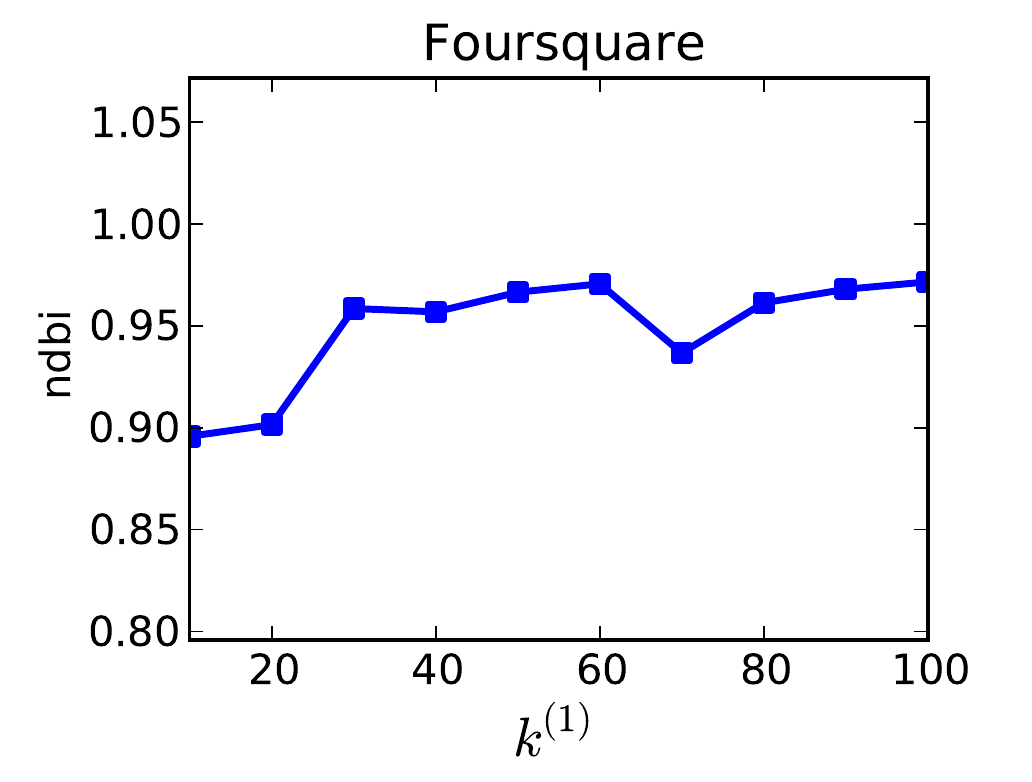}
    \end{minipage}
}
\subfigure[$k^{(1)}$-ndbi (Twitter)]{ \label{eg_fig9_2}
    \begin{minipage}[l]{0.45\columnwidth}
      \centering
      \includegraphics[width=1.0\textwidth]{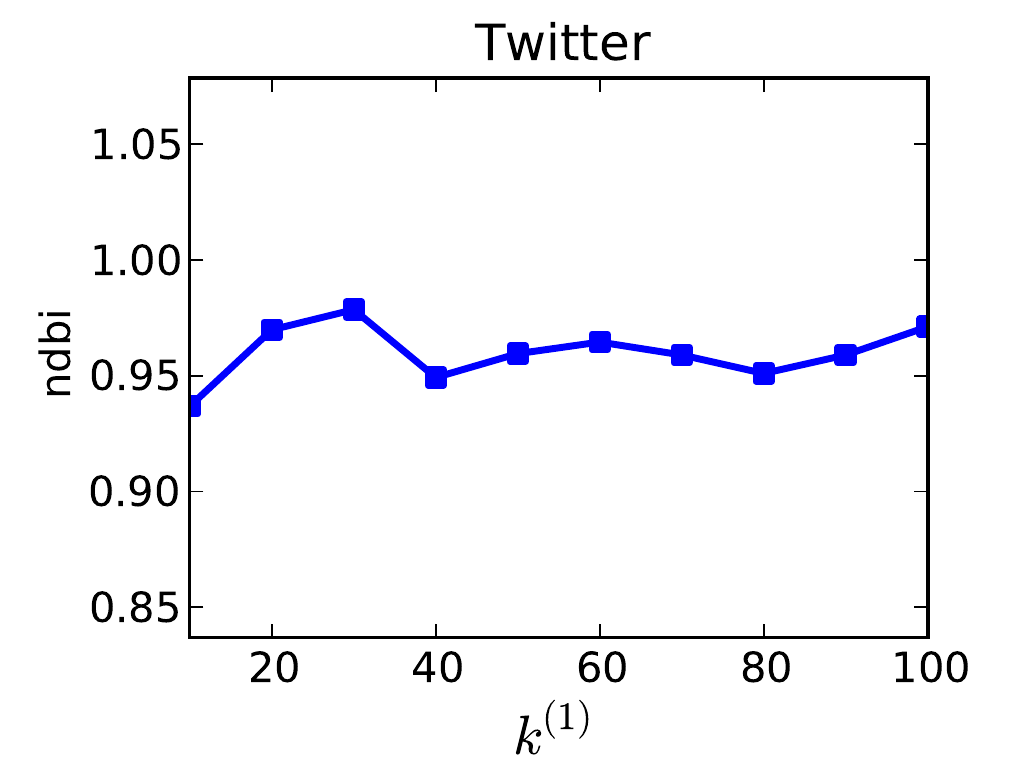}
    \end{minipage}
}
\subfigure[$k^{(1)}$-rand]{\label{eg_fig9_3}
    \begin{minipage}[l]{0.45\columnwidth}
      \centering
      \includegraphics[width=1.0\textwidth]{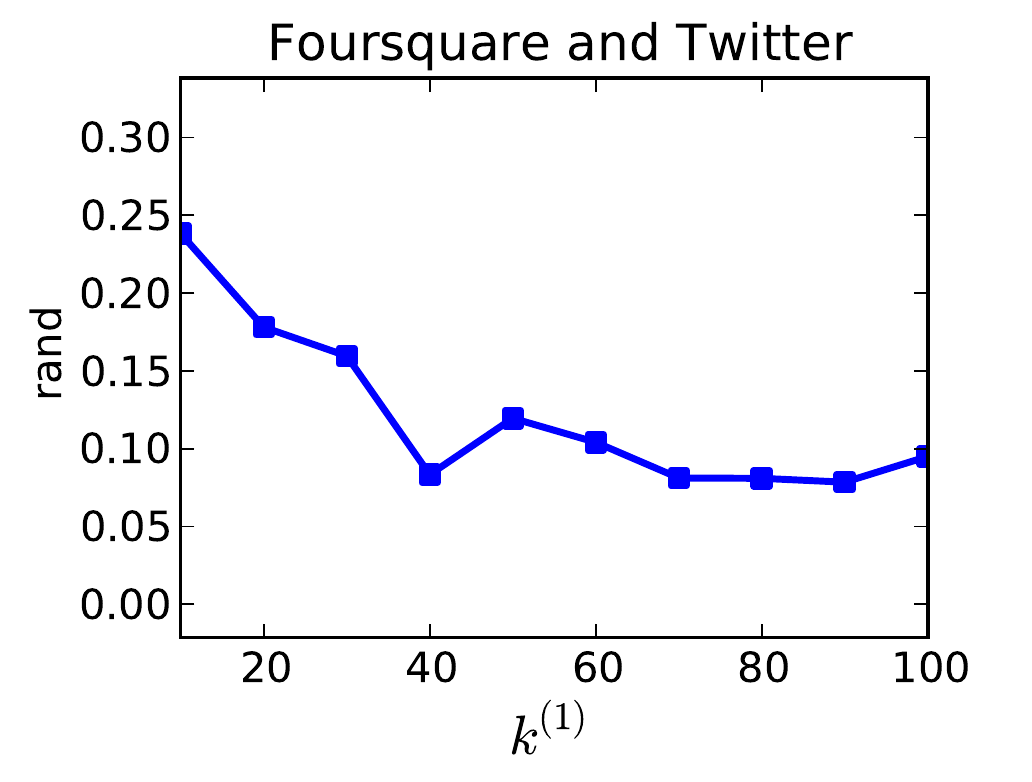}
    \end{minipage}
}
\subfigure[$k^{(1)}$-$IQC^{ndbi}_{rand}$]{\label{eg_fig9_4}
    \begin{minipage}[l]{0.45\columnwidth}
      \centering
      \includegraphics[width=1.0\textwidth]{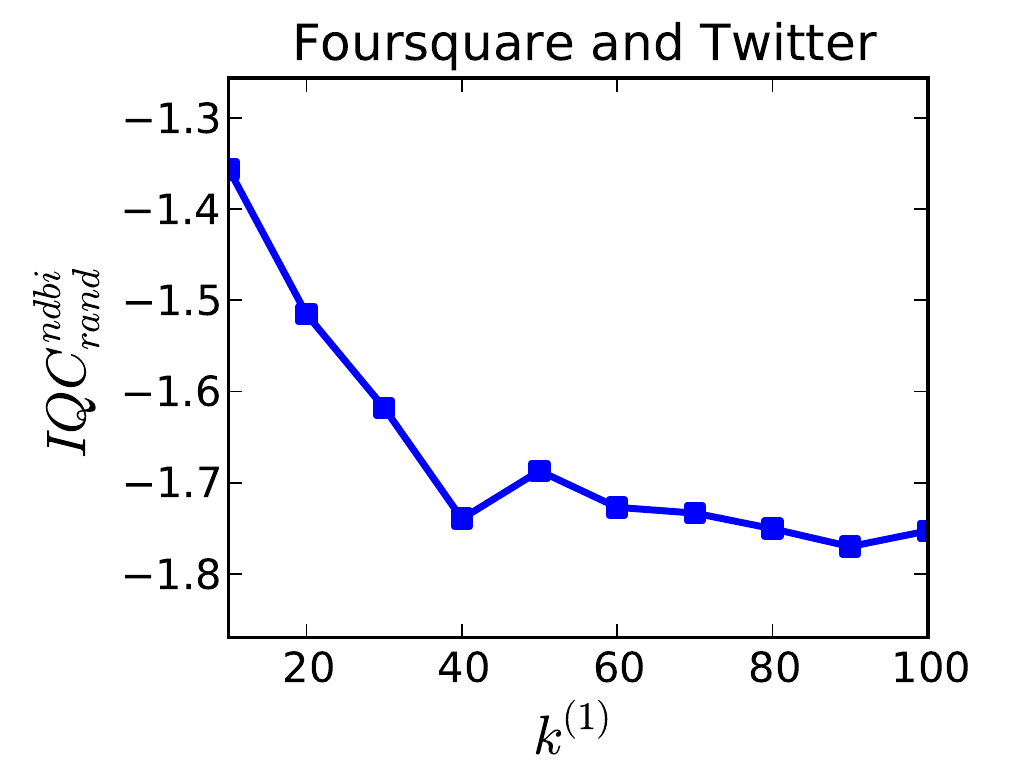}
    \end{minipage}
}

\subfigure[$k^{(2)}$-ndbi (Foursquare)]{\label{eg_fig10_1}
    \begin{minipage}[l]{0.45\columnwidth}
      \centering
      \includegraphics[width=1.0\textwidth]{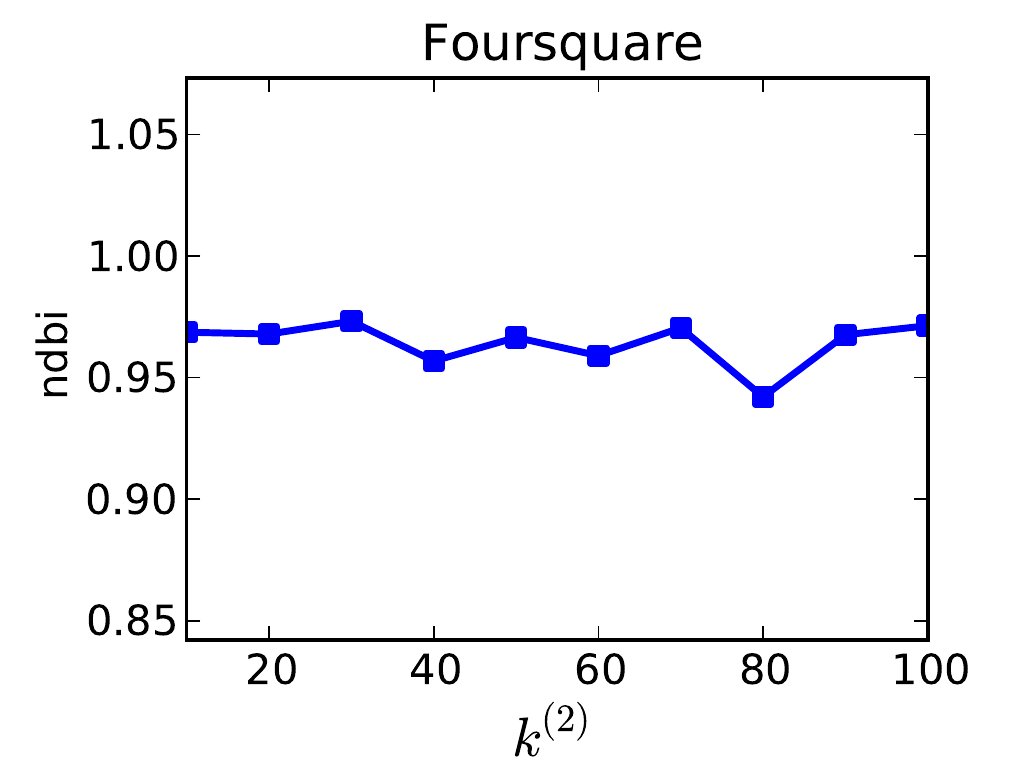}
    \end{minipage}
}
\subfigure[$k^{(2)}$-ndbi (Twitter)]{ \label{eg_fig10_2}
    \begin{minipage}[l]{0.45\columnwidth}
      \centering
      \includegraphics[width=1.0\textwidth]{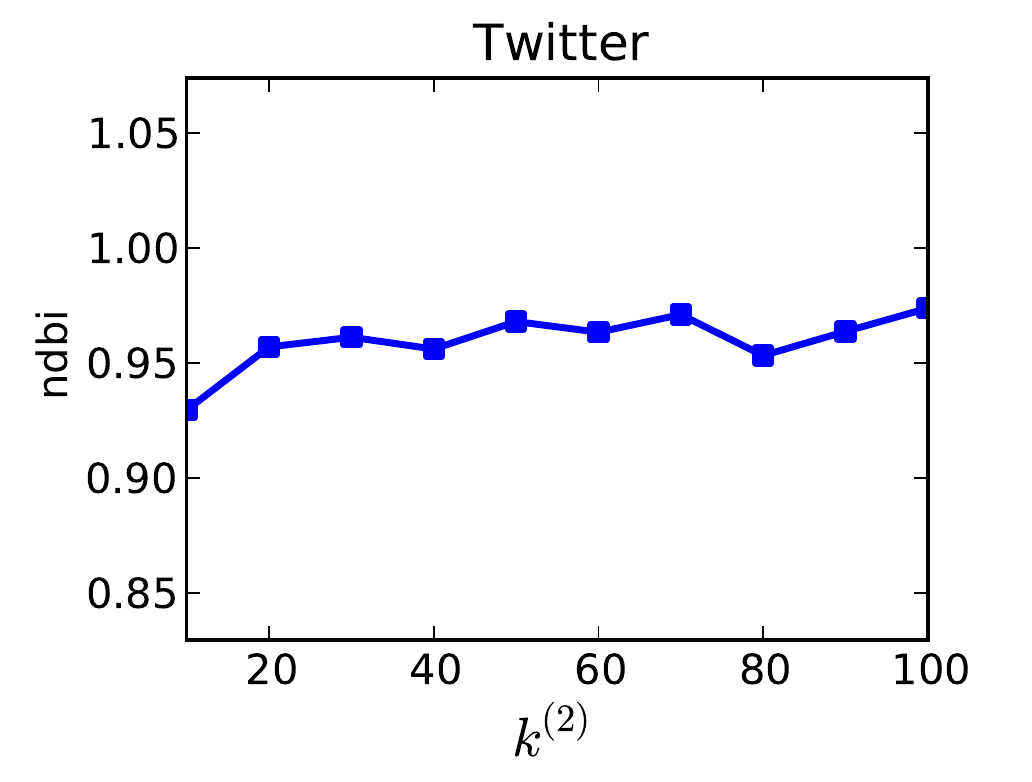}
    \end{minipage}
}
\subfigure[$k^{(2)}$-rand]{ \label{eg_fig10_3}
    \begin{minipage}[l]{0.45\columnwidth}
      \centering
      \includegraphics[width=1.0\textwidth]{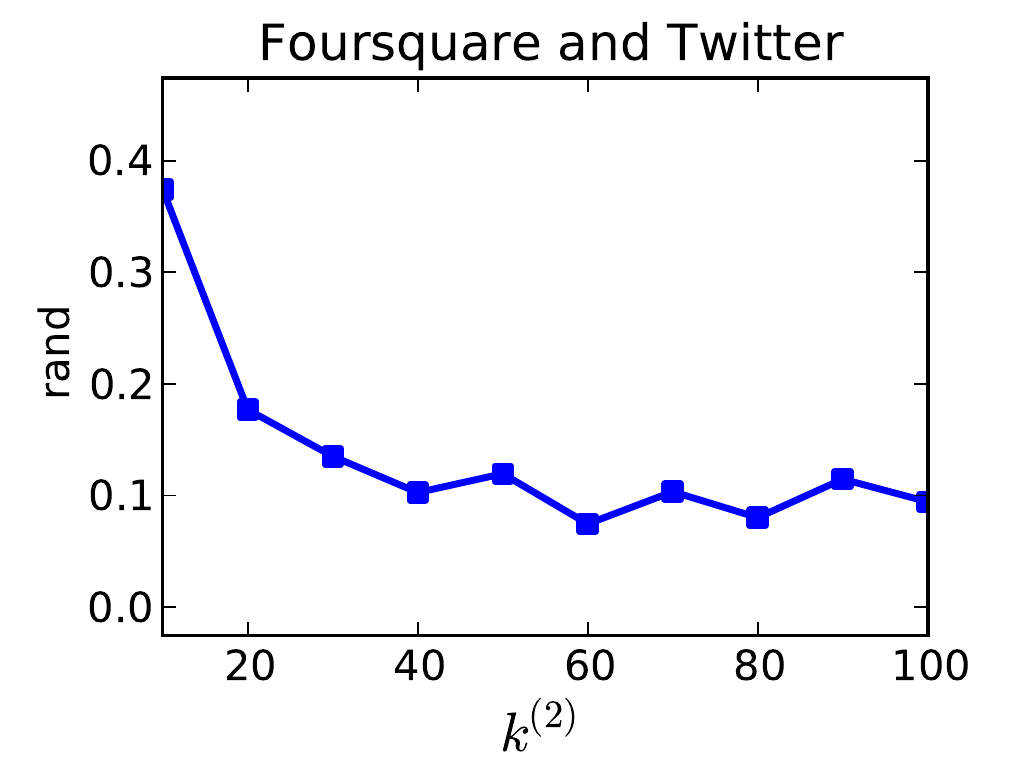}
    \end{minipage}
}
\subfigure[$k^{(2)}$-$IQC^{ndbi}_{rand}$]{\label{eg_fig10_4}
    \begin{minipage}[l]{0.45\columnwidth}
      \centering
      \includegraphics[width=1.0\textwidth]{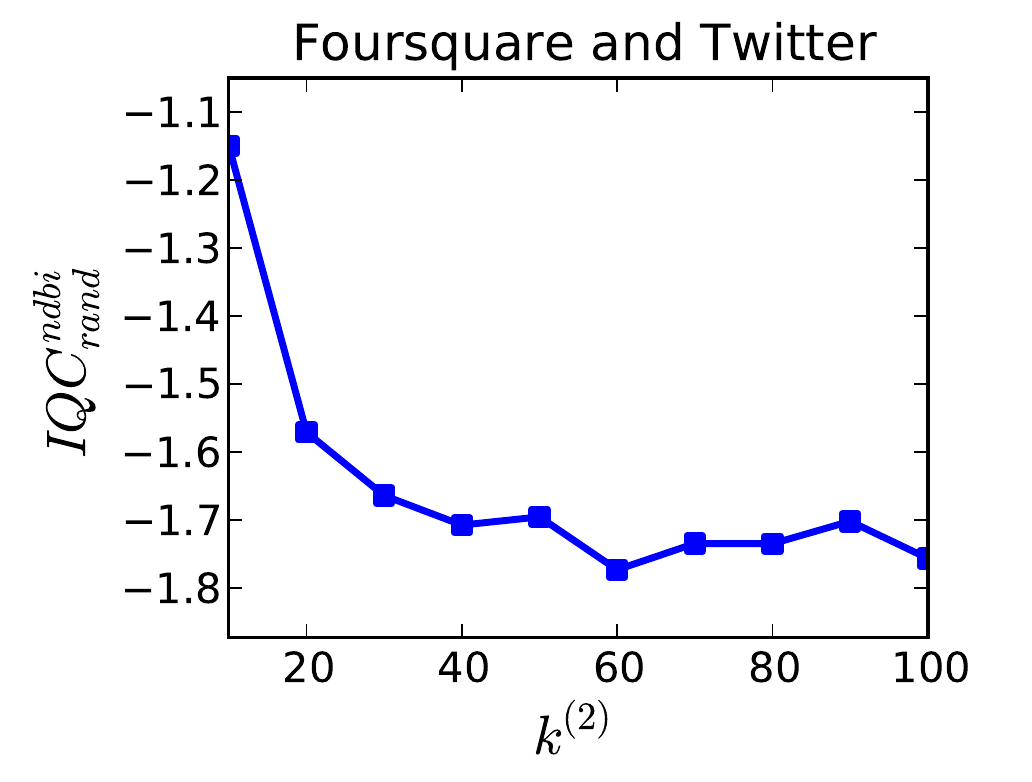}
    \end{minipage}
}
\caption{Analysis of parameters $k^{(1)}$ and $k^{(2)}$.}\label{eg_fig9}
\end{figure}

\subsection{Experiment Results}
The experiment results are available in Tables~\ref{tab:setting1}-\ref{tab:setting3}. To show the effects of the anchor links, we use the same networks but randomly sample a proportion of anchor links from the networks, whose number is controlled by $\sigma \in \{0.1, 0.2,\\ \cdots, 1.0\}$, where $\sigma = 0.1$ means that 10\% of all the anchor links are preserved and $\sigma = 1.0$ means that all the anchor links are preserved.

Table~\ref{tab:setting1} displays the clustering results of different methods in Foursquare and Twitter respectively under the evaluation of ndbi, entropy, density and silhouette. As shown in these two tables, {\our} can achieve the highest ndbi score in both Foursquare and Twitter for different sample rate of anchor links consistently. The entropy of the clustering results achieved by {\our} is the lowest among all other comparison methods and is about 70\% lower than {\ourinfluence}, 40\% lower than {\ourcut} and {\ourkmeans} in both Foursquare and Twitter. In each community detected by {\our}, the social connections are denser than that of {\ourinfluence} , {\ourcut} and {\ourkmeans}. Similar results can be obtained under the evaluation of silhouette, the silhouette score achieved by {\our} is the highest among all comparison methods. So, {\our} can achieve better results than modified multi-view and isolated clustering methods under the evaluation of \textit{quality metrics}.


Table~\ref{tab:setting3} shows the clustering results on the aligned networks under the evaluation of consensus metrics, which include rand, vi, nmi and mi. As shown in Table~\ref{tab:setting3}, {\our} can perform the best among all the comparison methods under the evaluation of consensus metrics. For example, the rand score of {\our} is the lowest among all other methods and when $\sigma = 0.5$, the rand score of {\our} is 20\% lower than {\ourinfluence}, 72\% lower than {\ourcut} and {\ourkmeans}. Similar results can be obtained for other evaluation metrics, like when $\sigma = 0.5$ , the vi score of {\our} is about half of the the score of {\ourinfluence}; the nmi and mi score of {\our} is the triple of that of{\ourkmeans}. As a result, {\our} can achieve better performance than both modified multi-view and isolated clustering methods under the evaluation of \textit{consensus metrics}.

Table~\ref{tab:setting4} is the clustering results of different methods evaluated by the $IQC$ metrics. As shown in Table~\ref{tab:setting4}, the $IQC^{ndbi}_{rand}$, $IQC^{ent.}_{vi}$, $IQC^{dens.}_{nmi}$, $IQC^{sil.}_{mi}$ scores of {\our} are all the lowest among all comparison methods. As mentioned above, lower $IQC$ score corresponds to better clustering results, {\our} can outperform all other baseline methods consistently under the evaluation of all $IQC$ metrics. In sum, {\our} can perform better than both modified multi-view and isolated clustering methods evaluated by $IQC$ metrics.

According to the results shown in Tables~\ref{tab:setting1}-\ref{tab:setting4}, we observe that the performance of {\our} doesn't varies much as $\sigma$ changes. The possible reason can be that, in method {\our}, normalized clustering discrepancy is applied to infer the clustering confidence matrices. As $\sigma$ increases in the experiments, more anchor links are added between networks, part of whose effects will be neutralized by the normalization of clustering discrepancy and doesn't affect the performance of {\our} much.


\subsection{Convergence Analysis}

{\our} can compute the solution of the optimization function with Curvilinear Search method, which can update matrices $\mb{X}^{(1)}$ and $\mb{X}^{(2)}$ alternatively. This process will continue until convergence. To check whether this process can stop or not, in this part, we will analyze the convergence of $\mb{X}^{(1)}$ and $\mb{X}^{(2)}$. In Figure~\ref{eg_fig8}, we show the $L^1$ norm of matrices $\mb{X}^{(1)}$ and $\mb{X}^{(2)}$, $\left\| \mb{X}^{(1)} \right\|_1$ and $\left\| \mb{X}^{(2)} \right\|_1$, in each iteration of the updating algorithm, where the $L^p$ norm of matrix $\mb{X}$ is $\left\| \mb{X} \right\|_p = (\sum_i \sum_i {X_{ij}}^p)^{\frac{1}{p}}$. As shown in Figures~\ref{eg_fig8}, both $\left\| \mb{X}^{(1)} \right\|_1$ and $\left\| \mb{X}^{(2)} \right\|_1$ can converge in less than $200$ iterations.


\begin{figure*}[t]
\centering
\subfigure[$\theta$-ndbi (Foursquare)]{\label{eg_fig11_1}
    \begin{minipage}[l]{0.45\columnwidth}
      \centering
      \includegraphics[width=1.0\textwidth]{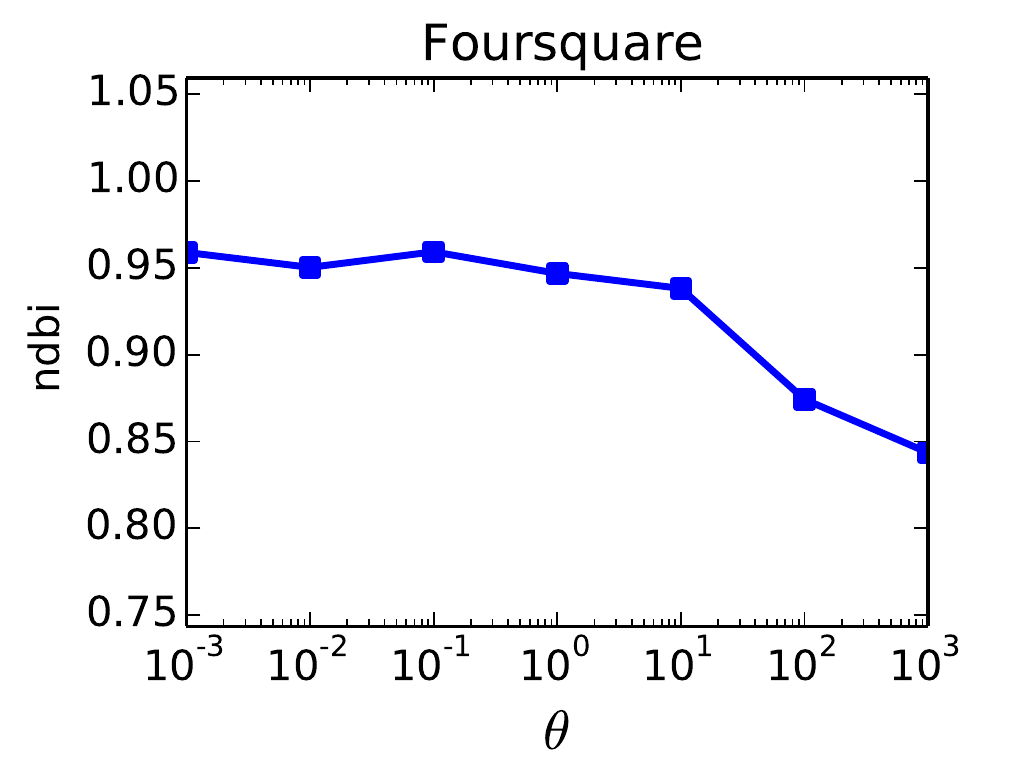}
    \end{minipage}
}
\subfigure[$\theta$-ndbi (Twitter)]{ \label{eg_fig11_2}
    \begin{minipage}[l]{0.45\columnwidth}
      \centering
      \includegraphics[width=1.0\textwidth]{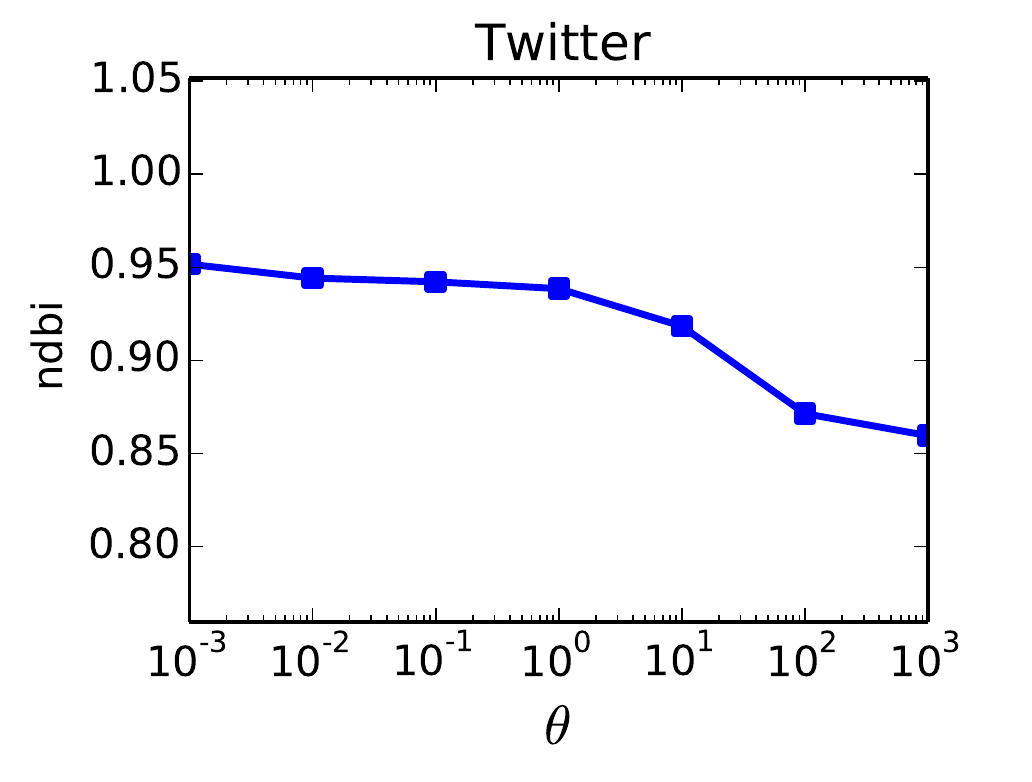}
    \end{minipage}
}
\subfigure[$\theta$-rand]{\label{eg_fig11_3}
    \begin{minipage}[l]{0.45\columnwidth}
      \centering
      \includegraphics[width=1.0\textwidth]{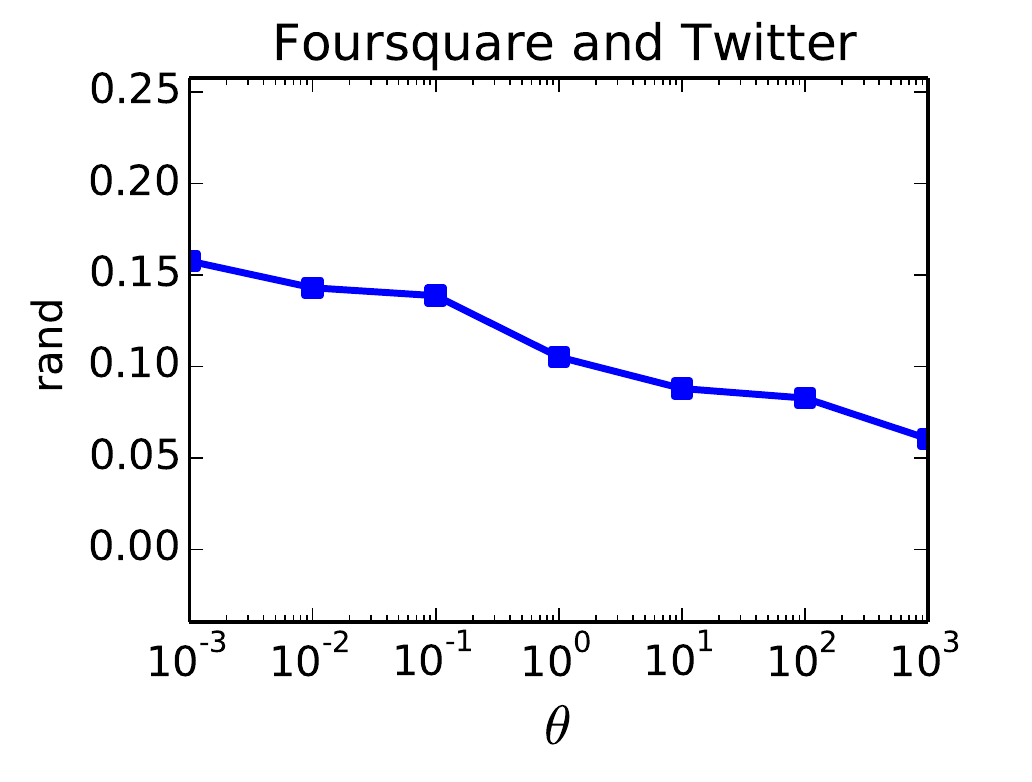}
    \end{minipage}
}
\subfigure[$IQC^{ndbi}_{rand}$ metric]{\label{eg_fig11_4}
    \begin{minipage}[l]{0.45\columnwidth}
      \centering
      \includegraphics[width=1.0\textwidth]{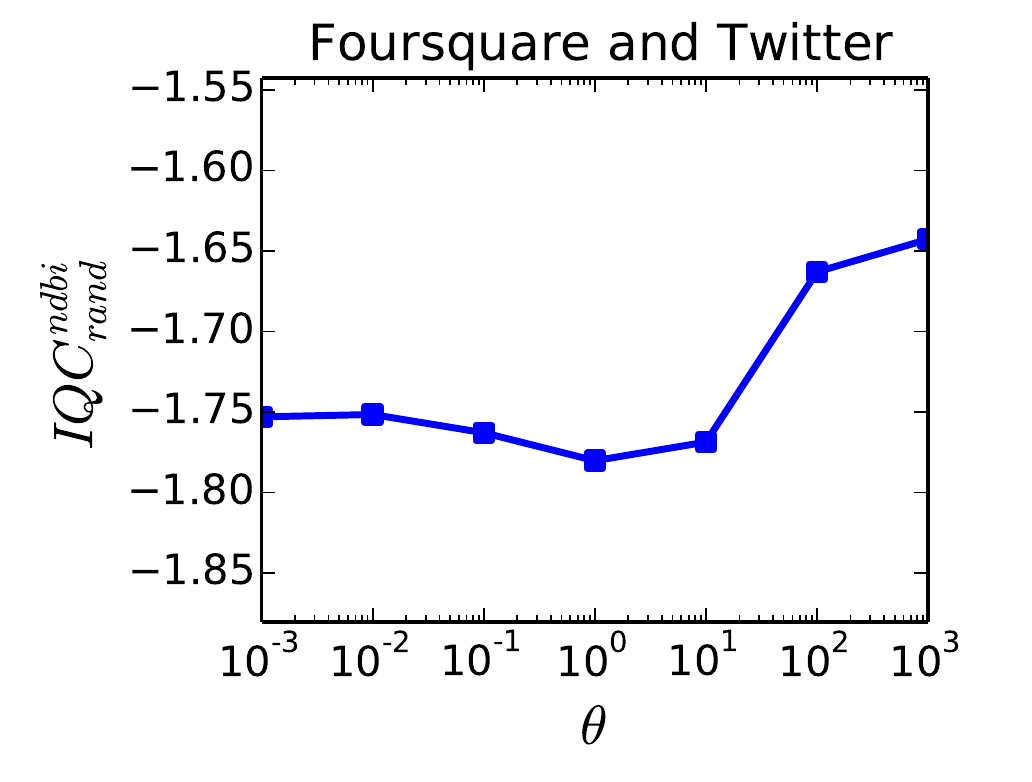}
    \end{minipage}
}
\caption{Analysis of parameter $\theta$.}\label{eg_fig11}
\end{figure*}

\subsection{Parameter Analysis}
In method {\our}, we have three parameters: $k^{(1)}$, $k^{(2)}$ and $\theta$, where $k^{(1)}$ and $k^{(2)}$ are the numbers of clusters in Foursquare and Twitter networks respectively, while $\theta$ is the weight of the normalized discrepancy term in the object function. In the pervious experiment, we set $k^{(1)} = 50$, $k^{(2)} = 50$ and $\theta = 1.0$. Here we will analyze the sensitivity of these parameters in details.

To analyze $k^{(1)}$, we fix $k^{(2)} = 50$ and $\theta = 1.0$ but assign $k^{(1)}$ with values in $\{10, 20, 30, 40, 50, 60, 70, 80, 90, 100\}$. The clustering results of {\our} with different $k^{(1)}$ evaluated by $ndbi$, $rand$ and $IQC^{ndbi}_{rand}$ metrics are given in Figures~\ref{eg_fig9_1}-\ref{eg_fig9_4}. As shown in the figures, the results achieved by {\our} are very stable for $k^{(1)}$ with in range $[40, 100]$ under the evaluation of $ndbi$ in both Foursquare and Twitter. Similar results can be obtained in Figures~\ref{eg_fig9_3}-\ref{eg_fig9_4}, where the performance of {\our} on aligned networks is not sensitive to the choice of $k^{(1)}$ for $k^{(1)}$ in range $[40, 100]$ under the evaluation of both $rand$ and $IQC_{ndbi, rand}$. In a similar way, we can study the sensitivity of parameter $k^{(2)}$, the results about which are shown in Figures~\ref{eg_fig10_1}-\ref{eg_fig10_4}.

 An interesting phenomenon is that the pre-defined number of clusters in the Foursquare network can also affect {\our}'s performance in the Twitter network. As shown in Figure~\ref{eg_fig9_2}, the performance of {\our} is the best in the Twitter network when $k^{(1)}$ is assigned with $30$, as the $ndbi$ score of {\our} is the highest when $k^{(1)} = 30$. Figures~\ref{eg_fig9_3}-~\ref{eg_fig9_4} show the performance of {\our} under the evaluation of $rand$ and $IQC_{ndbi, rand}$.  {\our} performs the best when $k^{(1)} = 40$ under the evaluation of the $rand$ metric and achieves the best performance when $k^{(1)} = 40$(or 90) evaluated by $IQC_{ndbi, rand}$.


To analyze the parameter $\theta$, we set both $k^{(1)}$ and $k^{(2)}$ as $50$ but assign $\theta$ with values in \{0.001, 0.01, 0.1, 1.0, 10.0, 100.0, 1000.0\}. The results are shown in Figure~\ref{eg_fig11}, where when $\theta$ is small, e.g., $0.001$, the $ndbi$ scores achieved by {\our} in both Foursquare and Twitter are high but the $rand$ score is not good ($rand$ is inversely proportional). On the other hand, large $\theta$ can lead to good $rand$ score but bad $ndbi$ scores in both Foursquare and Twitter. As a result, (1) large $\theta$ prefers consensus results, (2) small $\theta$ can preserve network characteristics and prefers high quality results. Meanwhile, considering the clustering quality and consensus simultaneously, {\our} can achieve the best performance when $\theta = 1$, as the $IQC^{ndbi}_{rand}$ is the lowest when $\theta = 1$ in Figure~\ref{eg_fig11_4}.

\section{Related Work} \label{sec:relatedwork}

Clustering is a very broad research area, which include various types of clustering problems, e.g., consensus clustering \cite{LBRFFP13, LD13}, multi-view clustering \cite{BS04, CNH13}, multi-relational clustering \cite{YHY07}, co-training based clustering \cite{KD11}, and dozens of papers have been published on these topics. Lourenco et al. \cite{LBRFFP13} propose a probabilistic consensus clustering method by using evidence accumulation. Lock et al. propose a bayesian consensus clustering method in \cite{LD13}. Meanwhile, Bickel et al. \cite{BS04} propose to study the multi-view clustering problem, where the attributes of objects are split into two independent subsets. Cai et al. \cite{CNH13} propose to apply multi-view K-Means clustering methods to big data. Yin et al. \cite{YHY07} propose a user-guided multi-relational clustering method, CrossClus, to performs multi-relational clustering under user's guidance. Kumar et al. propose to address the multi-view clustering problem based on a co-training setting in \cite{KD11}. 

A multi-view clustering paper which is correlated to the problem studied in this paper is \cite{CZGW13}, which relaxes the \textit{one-to-one} constraint in traditional multi-view clustering problems to uncertain mappings. Weights of such mappings need to be decided by prior domain knowledge and each view is actually a homogeneous network. To regularize the clustering results, a cost function called \textit{clustering disagreement} is introduced in \cite{CZGW13}, whose absolute value of all nodes in multiple views is involved in the optimization. Different from \cite{CZGW13}: (1) the constraint on anchor links in this paper is \textit{one-to-one} and no domain knowledge is required, (2) each network involves different users and contains heterogeneous information, (3) we apply clustering discrepancy to constrain the community structures of anchor users only and non-anchor users are pruned before calculating discrepancy cost, and (4) the clustering discrepancy is normalized before being applied in mutual clustering objective function.

Clustering based community detection in online social networks is a hot research topic and many different techniques have been proposed to optimize certain measures of the results, e.g., modularity function \cite{NG04}, and normalized cut \cite{SM00}. Malliaros et al. give a comprehensive survey of correlated techniques used to detect communities in networks in \cite{MV13} and a detailed tutorial on spectral clustering has been given by Luxburg in \cite{L07}. These works are mostly studied based on homogeneous social networks. However, in the real-world online social networks, abundant heterogeneous information generated by users' online social activities exist in online social networks. Sun et al. \cite{SYH09} studies ranking-based clustering on heterogeneous networks, while Ji et al. \cite{JHD11} studies ranking-based classification problems on heterogeneous networks. Coscia et al. \cite{CGP11} proposes a classification based method for community detection in complex networks and Mucha et al. study the community structures in multiplex networks in \cite{MRMPO10}.

In recent years, researchers' attention has started to shift to study multiple heterogeneous social networks simultaneously. Kong et al. \cite{KZY13} are the first to propose the concepts of \textit{aligned networks} and \textit{anchor links}. Across aligned social networks, different social network application problems have been studied, which include different cross-network link prediction/transfer \cite{ZKY13, ZKY14, ZYZ14, ZY15_2}, emerging network clustering \cite{ZY15} and large-scale network community detection \cite{JZYYL14}, inter-network information diffusion and influence maximization \cite{ZZWYX15}.




\section{Conclusion}\label{sec:conclusion}
In this paper, we have studied the \textit{mutual clustering} problem across multiple partially aligned heterogeneous online social networks. A novel clustering method, {\our}, has been proposed to solve the \textit{mutual clustering} problem. We have proposed a new similarity measure, HNMP-Sim, based on social meta paths in the networks. {\our} can achieve very good clustering results in all aligned networks simultaneously with full considerations of network difference problem as well as the connections across networks. Extensive experiments conducted on two real-world partially aligned heterogeneous networks demonstrate that {\our} can perform very well in solving the \textit{mutual clustering} problem. 


\label{sec:appendix}
\section{Acknowledgement}
This work is supported in part by NSF through grants CNS-1115234 and OISE-1129076, Google Research Award, and the Pinnacle Lab at Singapore Management University.


\end{document}